\documentclass[preprint,12pt]{elsarticle}
\usepackage{amssymb}
\usepackage{amsmath}
\usepackage{slashed}
\usepackage{undertilde}
\usepackage{multirow}
\usepackage{color}
\usepackage{framed}
\usepackage{graphicx}

  \newcounter{bla}

\definecolor{lightgrey}{gray}{0.9}
 \def\btab#1\etab{\begin{tabular}{p{51mm}p{71mm}}#1\end{tabular}}
\def\btabx#1\etabx{\begin{tabular}{p{66mm}p{56mm}}#1\end{tabular}}
\def\btaby#1\etaby{\begin{tabular}{p{31mm}p{91mm}}#1\end{tabular}}
\def\btabz#1\etabz{\begin{tabular}{p{37mm}p{85mm}}#1\end{tabular}}
\def\btabshort#1\etabshort{\begin{tabular}{p{5mm}p{116mm}}#1\end{tabular}}
\def\btabwide#1\etabwide{\begin{tabular}{p{81mm}p{41mm}}#1\end{tabular}}
 \def\bcen{\begin{center}}
 \def\ecen{\end{center}}
 \def\bgfb#1\egfb{\bcen\fcolorbox{black}{lightgrey}{\parbox{132mm}{\btab#1\etab}}\ecen}
\def\bgfbx#1\egfbx{\bcen\fcolorbox{black}{lightgrey}{\parbox{132mm}{\btabx#1\etabx}}\ecen}
\def\bgfbalign#1\egfbalign{\bcen\fcolorbox{black}{lightgrey}{\parbox{132mm}{\btaby#1\etaby}}\ecen}
\def\bgfbalignstretch#1\egfbalignstretch{\bcen\fcolorbox{black}{lightgrey}{\parbox{132mm}{\btabz#1\etabz}}\ecen}
\def\bgfbalignshort#1\egfbalignshort{\bcen\fcolorbox{black}{lightgrey}{\parbox{132mm}{\btabshort#1\etabshort}}\ecen}
\def\bgfbwide#1\egfbwide{\bcen\fcolorbox{black}{lightgrey}{\parbox{130mm}{\btabwide#1\etabwide}}\ecen}

\newcommand{\ie}{{\it i.e.}}
\newcommand{\eg}{{\it e.g.}}

\newcommand{\hc}{{\rm h.c.}}

\newcommand{\feynrules}{{\sc FeynRules}}
\newcommand{\feynarts}{{\sc FeynArts}}
\newcommand{\mathematica}{{\sc Mathematica}}

\newcommand{\be}{\begin{equation}}
\newcommand{\nn}{\nonumber}
\newcommand{\ee}{\end{equation}}
\def\bsp#1\esp{\begin{split}#1\end{split}}
\def\bpm{\begin{pmatrix}}
\def\epm{\end{pmatrix}}

\newcommand{\del}{\partial}
\newcommand{\lag}{{\cal L}}

\newcommand{\e}{\varepsilon}
\newcommand{\alphadot}{{\dot\alpha}}
\newcommand{\betadot}{{\dot\beta}}

\journal{Computer Physics Communications}

\begin{document}

\begin{frontmatter}

  \title{A superspace module for the FeynRules package}
  \author[a]{Claude Duhr}
  \address[a]{Institute for Particle Physics Phenomenology, University of Durham,
     Durham, \mbox{DH1 3LE}, U.K.\\\textit{E-mail address:}claude.duhr@durham.ac.uk}
  \author[b]{Benjamin Fuks}
  \address[b]{Institut Pluridisciplinaire Hubert Curien/D\'epartement Recherches
     Subatomiques, Universit\'e de Strasbourg/CNRS-IN2P3, 23 Rue du Loess, F-67037
     Strasbourg, France\\\textit{E-mail address:}benjamin.fuks@ires.in2p3.fr}

  \begin{abstract}
  \begin{flushright}
    \vspace{-10cm}  IPHC-PHENO-11-02  \\
     IPPP/11/11, DCPT/11/22\vspace{10cm}
  \end{flushright}
    We describe an additional module for the \mathematica\ package
    \feynrules\ that allows for an easy building of any $N=1$ supersymmetric
    quantum field theory, directly in superspace. After the superfield content of a 
    specific model has been implemented, the user can study the properties of the model, 
    such as the supersymmetric transformation laws
    of the associated Lagrangian, directly in \mathematica. 
    While the model dependent parts of the latter,
    \ie, the soft supersymmetry-breaking Lagrangian and the superpotential, 
    have to be provided by the user, the model independent pieces, such as the
    gauge interaction terms, are derived automatically. Using the
    strengths of the \feynrules\ program, it is then possible to derive all
    the Feynman rules associated to the model and implement them
    in all the Feynman diagram calculators interfaced to \feynrules\ in a straightforward way.
  \end{abstract}
  

\end{frontmatter}

\newpage

\noindent {\bf PROGRAM SUMMARY}                                               \\
  \begin{small}
  {\em Manuscript Title:} A superspace module for the FeynRules package.      \\
  {\em Authors:} Claude Duhr, Benjamin Fuks.                                  \\
  {\em Program Title:}                                                        \\
  {\em Journal Reference:}                                                    \\
  {\em Catalogue identifier:}                                                 \\
  {\em Licensing provisions:} none.                                           \\
  {\em Programming language:} \mathematica.                                   \\
  {\em Computer:} Platforms on which Mathematica is available.                \\
  {\em Operating system:} Operating systems on which Mathematica is available.
  {\em Classification:} 11.1 General, High Energy Physics and Computing.      \\
  \phantom{Classification:} 11.6 Phenomenological and Empirical Models
                                 and Theories.                                \\
  {\em External routines/libraries:} \feynrules.                              \\
  {\em Nature of problem:} Study of the properties of $N=1$ supersymmetric
    field theories using the superfield formalism, derivation of the 
    associated Lagrangians.                                                   \\
  {\em Solution method:} We use the \feynrules\ package and define internally
    the $N=1$ superspace. Then, we implement a module allowing to:
    \begin{enumerate}
    \item Perform the Grassmann variable series expansion so that any
       superfield expression can be developed in terms of the component 
       fields. The resulting expression is thus suitable to be treated by the
       \feynrules\ package directly.                                          \\
    \item Execute a set of operations associated to the superspace, such as 
       the superderivatives of an expression or the calculation of its 
       supersymmetric transformation laws.
    \end{enumerate}
  {\em Restrictions:} Superfields related to spin 3/2 and 2 particles are not
    implemented.                                                              \\
  {\em Unusual features:} All calculations in the internal routines are
    performed completely. The only hardcoded core is the Grassmann variable 
    algebra.                                                                  \\
  {\em Running time:} It depends on the user's purposes. The extraction of a
    Lagrangian in terms of the component fields may take a few minutes 
    for a complete model with complex mixing between the fields.\\
\end{small}

\newpage

\section{Introduction}
\label{sec:intro}
Supersymmetric (SUSY) theories are among the most popular extensions of the Standard
Model (SM) of particle physics and widely studied both on the theoretical and experimental 
sides. Apart from relating bosons with
fermions and unifying internal and space-time symmetries, 
they address a set of conceptual problems of the SM, such as the large hierarchy between the
electroweak and the Planck scale, gauge coupling unification at high energy or
the dark matter in the universe. However, as the supersymmetric partners of
the SM particles have not yet been observed, supersymmetry must be broken at low energies, and 
in order to remain a viable solution to the hierarchy problem, this breaking must
be soft, predicting massive superpartners around the TeV scale. Therefore, the
quest for SUSY particles is one of the main goals of the present 
high-energy collider experiments such as the Tevatron at Fermilab or the Large Hadron Collider at CERN.

Studying the hadron-collider phenomenology of models  that go beyond the Standard Model 
requires the use of Monte Carlo event generators, based,
on the one hand, on a proper modeling of the strong interactions to describe correctly the
parton showering, fragmentation and hadronization, and, on the other hand, on
the calculation of matrix-elements underlying the hard-scattering process. 
This last step requires the
implementation of the complete set of Feynman rules associated to a given model, often one vertex at the time. This task, tedious if undertaken manually, has been rendered much easier with
the use of packages such as {\sc LanHep} \cite{Semenov:1998eb, Semenov:2008jy}
or \feynrules\ \cite{Christensen:2008py, Christensen:2009jx,
Christensen:2010wz}, starting from the Lagrangian of the theory and exporting
the corresponding Feynman rules to one or several Monte Carlo generators.
Specific tools dedicated to SUSY theories, such as the {\sc Sarah} package
\cite{Staub:2009bi, Staub:2010jh}, also focusing on the same issue, in addition to
the generation of the model mass spectrum at the one-loop level.
However, all of these tools only address the generation of the Lagrangian
(partially automatically or not) in the usual spacetime, and are not suitable
for computations in superspace.

Supersymmetric model building and the underlying calculations are often
technically long and painstaking. The latter can be rendered easier if working
within the so-called superspace formalism \cite{Salam:1974yz,Ferrara:1974ac}, a
natural framework for SUSY model building. In this paper, we present an
extension to the \feynrules\ package, based on \mathematica\footnote{
{\sc Mathematica} is a registered trademark of Wolfram Research Inc.}, which
provides an environment to build supersymmetric theories directly in
superspace. The core of our program is a set of basic functions allowing
to expand superfields, \ie, functions defined on the superspace, in term of
their component fields, the usual scalar, fermionic and
vector fields of particle physics, in an automated way. As a consequence, the
user has only to worry about superfield expressions, in general much simpler
than their component fields counterparts. In addition, the module
contains a set of predefined functions for the supercharge and the superderivative
operators that allow the user to study the properties of the
SUSY theory under consideration. 

Lagrangians for phenomenologically relevant supersymmetric theories can very often 
be expressed as a sum of four terms. 
The first two describe the kinetic terms for the chiral and vector
supermultiplets and are independent of the specific model under consideration,
since they are fixed entirely by supersymmetry and gauge invariance. For this
reason, we have included into our code a set of functions to generate these
pieces of the Lagrangian, for any model, in an automated way. Hence, the implementation of a SUSY theory 
in the \feynrules\ package consists only in the setting of the superfield content,
the model parameters and the gauge symmetries of the theory as an input,
together with the model dependent parts of the Lagrangian, \ie, the
SUSY-breaking Lagrangian and the superpotential. This opens thus the way to a serious phenomenological study of entire classes of models whose implementation into Monte Carlo event generators was considered too complicated or too tedious so far. 
The \feynrules\ package including our superspace extension can be found, together
with an up-to-date manual, at http://feynrules.phys.ucl.ac.be.
 
The outline of the paper is as follows: In Section~\ref{sec:FR} we briefly review the main functionalities 
of the \feynrules\ package and the structure of the \feynrules\ model files. In Section~\ref{sec:superspace} we introduce our implementation of the Grassmann and supersymmetry algebras into  
\mathematica, before turning to the more specific case of the implementation of
chiral and vector superfields in Section~\ref{sec:superfields}. The main part of
the package, the simplification of superspace expressions and the generation and
expansion of supersymmetric Lagrangians with the help of the superspace module,
is discussed in Sections \ref{sec:simpli}, \ref{sec:SF2Parts}
and~\ref{sec:lagrangians}. Finally, in Section~\ref{sec:mssm} we illustrate the
use of the package on the example of the implementation of the Minimal
Supersymmetric Standard Model (MSSM) into \feynrules\ in terms of superfields.

\section{The FeynRules package}
\label{sec:FR}

\feynrules\ is a \mathematica\ package that allows to derive Feynman rules
directly from a Lagrangian~\cite{Christensen:2008py}. The information that the
user needs to provide consists in the particle content and the parameters of the
model, together with the Lagrangian that describes the interactions among the
different particles.
The Feynman rules can then be obtained automatically and the interaction vertices can be
exported to various matrix-element generators by means of a set of
translation interfaces included in the package. Presently, interfaces to {\sc
CalcHep}/{\sc CompHep} \cite{Pukhov:1999gg,Boos:2004kh,Pukhov:2004ca}, {\sc
FeynArts}/{\sc FormCalc} \cite{Hahn:2000kx,Hahn:2009bf}, {\sc MadGraph}/{\sc MadEvent}
\cite{Stelzer:1994ta,Maltoni:2002qb,Alwall:2007st,Alwall:2008pm}, {\sc Sherpa}
\cite{Gleisberg:2008ta} and {\sc Whizard/Omega}
\cite{Moretti:2001zz,Kilian:2007gr} are available.
In the following we briefly describe the basic features of the package and the model files, and we give a very brief introduction on how to run the code in order to derive the interaction vertices. For more details on both the \feynrules\ package as well as the interfaces, we refer the reader to Refs.~\cite{Christensen:2008py, Christensen:2009jx, Butterworth:2010ym, FRwebpage}.

The \feynrules\ model definition is an extension of the \feynarts\ model file format and consists in the definitions of the particles, parameters and gauge groups that characterize the model and the Lagrangian.
Following the original \feynarts~convention, particles are grouped into classes describing ``multiplets'' having the same quantum numbers, but possibly different masses. Each particle class is defined in terms of a set of class properties, given as a \mathematica~replacement list. For example, a Weyl fermion $\chi$ could be written as
\begin{center}
\begin{verbatim}
  W[1] == { ClassName      -> chi,
            SelfConjugate  -> False,
            Chirality      -> Left,
            Indices        -> {Index[Colour]} }  .
\end{verbatim}
\end{center}
This defines a left-handed Weyl fermion ({\tt W}) represented by the symbol {\tt chi}. Note that the antiparticle is automatically declared and represented by the symbol {\tt chibar}. The field carries an additional index labelled {\tt Colour}, representing its gauge charge under the QCD gauge group. Additional information, like the mass and width of the particles, as well as the $U(1)$ quantum numbers carried by the fields can also be included. A complete description of the particle classes and properties can be found in the \feynrules\ manual.

A Lagrangian is not only defined by its particle content, but also by the local and global symmetries defining the model. \feynrules\ allows to define gauge group classes in a way similar to the particle classes. As an example, the definition of the QCD gauge group can be written
\begin{center}
\begin{verbatim}
SU3C == { Abelian              -> False,
          GaugeBoson           -> G,
          CouplingConstant     -> gs,
          StructureConstant    -> f,
          Representations      -> {T, Colour} }  ,
\end{verbatim}
\end{center}
where the gluon field {\tt G} is defined together with the other fields during
the particle declaration. The declaration of abelian gauge groups is analogous.
\feynrules\ uses this information to construct the covariant derivative and
field strength tensor which the user can use in the Lagrangian.

The third main ingredient to define a model is the set of parameters which it depends upon. The declaration of the parameters follows the same lines as the declarations of the particle and gauge group classes. However, since the declaration of the parameters is not needed in order to understand how the superfield module works, we do not review it here but refer the reader to the \feynrules\ manual~\cite{Christensen:2008py}.

After having loaded the \feynrules\ package into \mathematica, the user can load the model and the model restrictions via the commands
\be
\texttt{LoadModel[ file1, file2, ... ]}\ ,\nn
\ee
where the model can be implemented in as many files as convenient or it can be implemented directly in the \mathematica~notebook in which case the list of files would be empty. The Lagrangian can now be entered directly into the notebook\footnote{Alternatively, the Lagrangian can also be included in the model file, in which case it is directly loaded together with the model file.} using standard \mathematica~commands, augmented by some special symbols representing specific objects like Dirac matrices. As an example, we show the Lagrangian,
\be\bsp
&\texttt{L = -1/4 FS[G, mu, nu, a] FS[G, mu, nu, a]}\\
&\, \,\quad\texttt{ + I chibar~.~sibar[mu]~.~DC[chi, mu]}\ ,\nn
\esp\ee
where {\tt FS[G, mu, nu, a]} and {\tt DC[chi, mu]} denote the $SU(3)_C$ field
strength tensors and covariant derivatives automatically defined by \feynrules,
respectively. At this stage, the user can perform a set of basic checks on the
Lagrangian (hermiticity, normalization of kinetic terms, \ldots), or directly
proceed to the derivation of the Feynman rules via the command
\be
\texttt{verts = FeynmanRules[ L ]}\ .\nn
\ee
\feynrules\ then computes all the interaction vertices associated with the Lagrangian {\tt L} and stores them in the variable {\tt verts}. The vertices can be used for further computations within \mathematica, or they can be exported to one of the various matrix-element generators for further phenomenological studies of the model.
The translation interfaces can be directly called from within the notebook, \eg, for the \feynarts~interface,
\be
\texttt{WriteFeynArtsOutput[ L ]}\ ,\nn
\ee
This will produce a file formatted for use in \feynarts.  All other interfaces are called in a similar way. 

The {\tt FeynmanRules} function, as well as the interfaces that call it, require
the Lagrangian to be written in four-dimensional space time. For supersymmetric theories, however, the most natural and most convenient way to write a Lagrangian is in terms of superfields, \emph{i.e.}, in terms of the explicit supermultiplets of the supersymmetry algebra. For this reason, we have extended the \feynrules\ package by a superspace module, which allows to write a Lagrangian in terms of superfields and to transform the latter into a four-dimensional Lagrangian involving component fields only, hence expressing the superfield Lagrangian in a way that can be processed immediately by the {\tt FeynmanRules} function. This new module will be described in detail in the next sections.

\section{Superspace and supersymmetry algebra}
\label{sec:superspace}

\subsection{Grassmann variables and fermionic fields}
Supersymmetric theories are naturally formulated in superspace 
\cite{Salam:1974yz,Ferrara:1974ac}, an extension
of the ordinary spacetime, defined, for $N=1$ supersymmetric theories by
 adjoining a Majorana spinor $(\theta_\alpha,\bar\theta^\alphadot)$ to
the usual spacetime coordinates $x^\mu$, where $\theta$ and $\bar\theta$ are
(Grassmannian)
two-component Weyl fermions. Using the recent implementation of Weyl
fermions into \feynrules\ \cite{Butterworth:2010ym}, the $\theta$ 
variables are hardcoded in the superspace module following the example of
Section \ref{sec:FR}, as
\begin{verbatim}
   W[x1000] == {
       TeX           -> \[Theta], 
       ClassName     -> theta, 
       Chirality     -> Left, 
       SelfConjugate -> False, 
       Unphysical    -> True}  .
\end{verbatim}
This field and its right-handed counterpart $\bar\theta$ can be accessed, after
loading 
\feynrules\ in \mathematica, through the symbols {\tt theta} and {\tt
thetabar}. By convention, the spin indices are assumed to be
lowered for both the $\theta$ and $\bar\theta$ variables,
\be
  \texttt{theta[alpha]} \leftrightarrow \theta_\alpha 
  \quad \text{and} \quad
  \texttt{thetabar[alphadot]} \leftrightarrow \bar\theta_{\dot\alpha} \ ,
\nn \ee
where we relate undotted and dotted indices to left and right-handed
fermions. Spin indices can be raised and lowered using the
rank-two antisymmetric tensors $\e_{\alpha\beta}$ and $\e_{\alphadot\betadot}$
\be
  \theta^\alpha = \e^{\alpha\beta}\theta_{\beta} \ , \quad 
  \theta_\alpha = \e_{\alpha\beta}\theta^{\beta} \ , \quad 
  \bar\theta^\alphadot = \e^{\alphadot\betadot}\bar\theta_{\betadot} 
    \quad \text{and} \quad 
  \bar\theta_\alphadot = \e_{\alphadot\betadot}\bar\theta^{\betadot} \ ,
\ee
with  $\e_{12}= 1$, $\e^{12}=-1$, $\e_{\dot 1 \dot 2}=1$ and $\e^{\dot 1 \dot
2}=-1$. This strictly follows the conventions of  Ref.\
\cite{FuksRausch}, and so
does the complete superspace module of \feynrules.
Regardless of the left or right-handed nature of the spin indices, the
tensors with lower and upper indices are implemented into \feynrules\ as {\tt
Deps} and {\tt Ueps}, respectively. An example is in order,
\be \bsp
 &\ \texttt{Ueps[alphadot,betadot] thetabar[betadot]} \leftrightarrow
   \e^{\alphadot\betadot}\bar\theta_{\betadot} = \bar\theta^\alphadot  \ , \\ 
 &\ \texttt{Deps[alpha,beta] Ueps[beta,gam] theta[gam]} 
    \leftrightarrow \e_{\alpha\beta} \e^{\beta\gamma} \theta_\gamma =
    \theta_\alpha \ .
\esp \nn \ee
The Levi-Civita tensors allow us to fix our conventions with respect to the summation on
the spin indices, and we define the dot product of two Weyl spinors as
\be
  \lambda\cdot\lambda^\prime = \lambda^\alpha\lambda^\prime_\alpha = 
    \e^{\alpha\beta} \lambda_\beta \lambda^\prime_\alpha \ , 
  \quad \text{and} \quad
  \bar\chi\cdot\bar\chi^\prime = \bar\chi_\alphadot\bar\chi^\prime{}^\alphadot =
      \e^{\alphadot\betadot} \bar\chi_\alphadot\bar\chi^\prime{}_\betadot\ ,
\label{eq:scalprod}\ee
where we have introduced some generic left and right-handed spinors
$\lambda^{(\prime)}$ and $\bar\chi^{(\prime)}$, respectively.

Computations in superspace require to keep
track not only of the position of the different spin indices, but also of the
ordering of the fermions.
The \feynrules\ superspace module addresses this issue in the following way.
First, any explicit spin index
will be considered, by default, as lowered. Second, we
have implemented the environment {\tt nc[chain]}, where {\tt chain} stands
for an ordered sequence of fermions with lower spin indices. 
As simple examples, scalar products such as in Eq.\
\eqref{eq:scalprod} are implemented as
\be\bsp
 &\  \lambda\cdot\lambda^\prime = 
     \e^{\beta\alpha} \lambda_\alpha\lambda^\prime_\beta \leftrightarrow 
     \texttt{nc[lambda[a], lambdaprime[b]] Ueps[b,a]} \ , \\
 &\  \bar\chi\cdot\bar\chi^\prime =
     \e^{\betadot\alphadot} \bar\chi_\betadot\bar\chi^\prime{}_\alphadot 
     \leftrightarrow 
     \texttt{nc[chi[bd], chiprime[ad]] Ueps[bd,ad]} \ .
     \label{eq:spin_convention}
\esp\ee

\subsection{Supercharges and superderivatives}
In $N=1$ supersymmetric theories, there is a single spinorial SUSY generator,
transforming as a Majorana
spinor $(Q_\alpha, \bar Q^\alphadot)$ under Lorentz transformations. 
Any translation in superspace can then be parameterized as
\be
  G(x,\theta,\bar\theta) = e^{i \big[x^\mu P_\mu + \theta \cdot Q + \bar Q \cdot
   \bar\theta\big]} \ ,
\ee
where $P_\mu$ denotes the generator of space-time translations.
A generic supersymmetric transformation then corresponds to $G(0, \epsilon,
\bar\epsilon)$, where $(\epsilon^\alpha, \bar\epsilon_\alphadot)$ denotes a
(spinorial) transformation parameter. Applying this
transformation to a generic superspace point,
$G(0,\epsilon,\bar\epsilon) G(x,\theta,\bar\theta)$, we can calculate the
variations of the spacetime coordinates $x$ and of the Grassmann
variables $\theta$ and $\bar\theta$ using the Baker-Campbell-Hausdorff
formula and the (anti)com\-mutation relations among the generators. 
Comparing with a direct
application of the supersymmetric generator $i (\epsilon \cdot Q + \bar Q \cdot
\bar \epsilon)$ on the coordinates $x$, $\theta$ and $\bar\theta$, we obtain
the action of the generators as differential operators acting on functions on superspace.
Similarly, starting from an action from the right, 
$G(x,\theta,\bar\theta)G(0,\epsilon,\bar\epsilon)$, we
obtain the expressions of the superderivatives $D_\alpha$ and $\bar
D_\alphadot$. In our conventions \cite{FuksRausch}, these four quantities are
given by
\be \bsp
  Q_\alpha = -i\Big( \frac{\del}{\del\theta^\alpha} + i \sigma^\mu{}_{\alpha
    \alphadot} \bar\theta^\alphadot \del_\mu \Big)\  , &\quad
  \bar Q_\alphadot = i\Big( \frac{\del}{\del \bar\theta^\alphadot} + i
    \theta^\alpha \sigma^\mu{}_{\alpha\alphadot}\del_\mu \Big) \ , \\
  D_\alpha = \frac{\del}{\del\theta^\alpha} - i \sigma^\mu{}_{\alpha\alphadot}
    \bar \theta^\alphadot \del_\mu\ , &\quad
  \bar D_\alphadot =  \frac{\del}{\del \bar\theta^\alphadot} - i\theta^\alpha
    \sigma^\mu{}_{\alpha\alphadot}\del_\mu \ .
\esp \label{eq:superch-and-der} \ee
The matrices $\sigma^\mu$, and for later
reference $\bar\sigma^\mu$, are the four-vectors corresponding to the usual Pauli
matrices.
Note that, since the actions from the left and from the right commute, the
supercharges and the superderivatives must anticommute. 

The generators $Q_\alpha$ and $\bar Q_\alphadot$ of the supersymmetric transformations, as well as the superderivatives $D_\alpha$ and $\bar D_\alphadot$, are called in \feynrules\ via the functions,
\be \bsp
 &\ \texttt{QSUSY[expression,alpha]} \leftrightarrow Q_\alpha
    (\text{expression})  \ , \\ 
 &\ \texttt{QSUSYBar[expression,alphadot]} \leftrightarrow \bar Q_\alphadot
    (\text{expression})  \ , \\ 
 &\ \texttt{DSUSY[expression,alpha]} \leftrightarrow D_\alpha
    (\text{expression})  \ , \\ 
 &\ \texttt{DSUSYBar[expression,alphadot]} \leftrightarrow \bar D_\alphadot
    (\text{expression})  \ .
\esp \nn \ee
It is assumed that all the indices appearing in
{\tt expression} must be written explicitly, using the environment {\tt nc}
presented above in order to keep track of the correct fermion ordering. The only
exception corresponds to the case of a single fermion, the ordering 
of the fermions being trivially irrelevant, and the {\tt nc} environment can be omitted. 
Hence, both
\be
 \texttt{DSUSY[chi[alpha], beta]}
 \quad \text{and}\quad
 \texttt{DSUSY[nc[chi[alpha]], beta]} 
\nn\ee
give a correct answer, whilst for two fermions, the correct synthax is 
\be
 \texttt{DSUSY[nc[chi[alpha],lambda[beta],gamma]} \ ,
\nn\ee
and not
\be
 \texttt{DSUSY[chi[alpha] lambda[beta],gamma]} \ .
\nn\ee

\renewcommand{\arraystretch}{1.5}
\begin{table}
\bgfbalign
\multicolumn{2}{c}{\textbf{Table~\ref{tab:superspace}: Superspace functionalities}}\\
{\tt theta[al]} & The Grassmann variable $\theta_\alpha$, where {\tt al}
  denotes the carried spin index. \\
{\tt thetabar[aldot]} & The Grassmann variable $\bar\theta_\alphadot$, where
  {\tt aldot} denotes the carried spin index. \\
{\tt Ueps[i,j]} & The rank-two antisymmetric tensor $\epsilon^{ij}$. \\
{\tt Deps[i,j]} & The rank-two antisymmetric tensor $\epsilon_{ij}$. \\
{\tt nc[seq] } & Ordered sequence of fermionic field(s), labelled by {\tt
  seq}, where each field carries explicitly its indices.\\
{\tt QSUSY[exp,al] } & Calculates the action of the supercharge $Q_\alpha$ on the
  expression {\tt exp}, where all indices must be written explicitly, using the
  {\tt nc} environment if relevant. The symbol {\tt al} is related to the spin
  index of the supercharge.\\
{\tt QSUSYBar[exp,ad] } & Calculates the action of the supercharge $\bar
  Q_\alphadot$ on the expression {\tt exp}, where all indices must be written
  explicitly, using the {\tt nc} environment if relevant. The symbol {\tt ad} is
  related to the spin index of the supercharge.\\
{\tt DSUSY[exp,al] } & Same as {\tt QSUSY}, but for the superderivative
  $D_\alpha$.\\
{\tt DSUSYBar[exp,ad] } & Same as {\tt QSUSYBar}, but for the superderivative
$\bar D_\alphadot$.\\
\egfbalign
\textcolor{white}{\caption{\label{tab:superspace}}}
\end{table}
\renewcommand{\arraystretch}{1}

\section{Implementing superfields into \feynrules}
\label{sec:superfields}

\subsection{Generic superfields}
Any function $\Phi(x,\theta,\bar\theta)$ defined on $N=1$ 
superspace is called a superfield and can be expanded into a Taylor series
with respect to the anticommuting coordinates $\theta$ and $\bar\theta$. Since
the square of an anticommuting object vanishes, the series has only a finite number
of terms and the most general expression for a scalar superfield is\footnote{The extension to non-scalar superfields is immediate.}
\be\label{eq:genSFa}\bsp
  \Phi(x,\theta,\bar\theta) = &\ z(x) 
    + \theta \!\cdot\! \xi(x) + \bar \theta \!\cdot\! \bar \zeta(x) 
    + \theta \!\cdot\! \theta f(x) + \bar \theta \!\cdot\! \bar \theta g(x) \\
 &\ + \theta \sigma^\mu \bar \theta V_\mu(x) 
    + \bar \theta \!\cdot\! \bar \theta\ \theta \!\cdot\! \omega(x) 
    + \theta \!\cdot\! \theta\ \bar \theta \!\cdot\! \bar \rho(x) 
    + \theta \!\cdot\! \theta\ \bar \theta \!\cdot\! \bar \theta d(x) \ , 
\esp \ee
where the various coefficients of the expansion form a so-called supermultiplet and are referred to as  the component fields
of the superfield. They correspond to the usual
scalar, fermionic and vector degrees of freedom used in particle physics. 
$z$,
$f$, $g$ and $d$ denote complex scalar
fields, whilst $\xi$, $\zeta$, $\omega$ and $\rho$ denote complex Weyl fermions and
$V_\mu$ is a complex vector field, leaving a total of 16 bosonic and 16
fermionic degrees of freedom. 

Using the {\tt nc} environment, the generic scalar superfield of Eq.~(\ref{eq:genSFa}) can be implemented into \feynrules\ in a straightforward way as
\be\bsp
  &\ \texttt{Phi} \texttt{ = }\\
  &\ \ \begin{array}{ll} 
    \texttt{z + } & \leftrightarrow z \\
    \texttt{nc[theta[sp], xi[sp2]] Ueps[sp2,sp] +} & \leftrightarrow \theta
      \!\cdot\! \xi\\
    \texttt{nc[thetabar[spd], zetabar[spd2]]} & \\
    \qquad \texttt{Ueps[spd,spd2] +} & \leftrightarrow
      \bar \theta \!\cdot\! \bar \zeta\\
    \texttt{nc[theta[sp], theta[sp2]] Ueps[sp2,sp] f +} & \leftrightarrow \theta
       \!\cdot\! \theta f \\
    \texttt{nc[thetabar[spd], thetabar[spd2]]} & \\
    \qquad \texttt{Ueps[spd,spd2] g +} &
       \leftrightarrow \bar \theta \!\cdot\! \bar \theta g\\
    \texttt{nc[theta[sp], thetabar[spd]] Ueps[sp2,sp]} & \\
    \qquad \texttt{Ueps[spd2,spd] si[mu,sp2,spd2] V[mu] +} & \leftrightarrow
      \theta \sigma^\mu \bar \theta V_\mu \\
    \texttt{nc[thetabar[spd], thetabar[spd2]]} & \\
    \qquad \texttt{Ueps[spd,spd2] Ueps[sp2,sp]} & \\
    \qquad \texttt{nc[theta[sp], omega[sp2]] +}& \leftrightarrow
      \bar \theta \!\cdot\! \bar \theta\ \theta \!\cdot\! \omega \\
    \texttt{nc[theta[sp], theta[sp2]] Ueps[sp2,sp]} &\\
    \qquad \texttt{nc[thetabar[spd], rhobar[spd2]]} &\\
    \qquad \texttt{Ueps[spd,spd2] +} &
    \leftrightarrow \theta \!\cdot\! \theta\ \bar \theta \!\cdot\! \bar \rho \\
    \texttt{nc[theta[sp], theta[sp2]] Ueps[sp2,sp]} & \\
    \qquad \texttt{nc[thetabar[spd], thetabar[spd2]]} &\\
    \qquad \texttt{Ueps[spd,spd2] d}
    &\leftrightarrow \theta \!\cdot\! \theta\ \bar \theta \!\cdot\! \bar \theta
      d \ ,
   \end{array}
\esp\label{eq:genSF}\ee
where we assume that the component fields have been defined properly in the
\feynrules\ model file (See Section~\ref{sec:FR}). In
Eq.~(\ref{eq:genSF}) we have made use of the object {\tt si[mu,alpha,alphadot]},
corresponding to the Pauli matrices $\sigma^\mu{}_{\alpha\dot\alpha}$. Note that
the appearance of the Levi-Civita tensor {\tt Ueps} is due to our convention
that all spin indices carried by {\it fermionic fields} are lowered. Similarly,
we could have used the matrices {\tt sibar[mu,alphadot,alpha]} defined with two
upper spin indices, $\bar\sigma^{\mu\alphadot\alpha}$, since 
\be
  \theta_\beta \epsilon^{\alpha\beta} \sigma^\mu{}_{\alpha\alphadot}
    \epsilon^{\alphadot\betadot} \bar \theta_\betadot = - \bar\theta_\alphadot
    \bar\sigma^{\mu\alphadot\alpha} \theta_\alpha \ .
\ee
The second option would have allowed to get an expression free from any
$\epsilon$-tensor.

Even though Eq.~\eqref{eq:genSF} is the canonical form in which superfields 
are represented inside the code, the form of Eq.\ ~\eqref{eq:genSF} can however be very painful 
to use in practice. For this reason, the package contains a function denoted {\tt ncc}, which has the 
same effect as {\tt nc}, but all spin indices and $\epsilon$-tensors are suppressed. 
The code then assumes that, within a given {\tt ncc} environment, all the suppressed indices are contracted 
according to the convention of Eq.\ \eqref{eq:spin_convention}, and outputs the
result into the canonical form. 
As an example, using the {\tt ncc} environment, 
Eq.~\eqref{eq:genSF} can be written in the more compact form,
\be\bsp
  &\ \texttt{Phi} \texttt{ = }\\
  &\ \ \begin{array}{ll} 
    \texttt{z + } & \leftrightarrow z \\
    \texttt{ncc[theta, xi]  +} & \leftrightarrow \theta
      \!\cdot\! \xi\\
    \texttt{ncc[thetabar, zetabar] +} & \leftrightarrow
      \bar \theta \!\cdot\! \bar \zeta\\
    \texttt{ncc[theta, theta]\, f +} & \leftrightarrow \theta
       \!\cdot\! \theta f \\
    \texttt{ncc[thetabar, thetabar] \,g +} &
       \leftrightarrow \bar \theta \!\cdot\! \bar \theta g\\
    \texttt{ncc[theta, si[mu], thetabar]  V[mu] +} & \leftrightarrow
      \theta \sigma^\mu \bar \theta V_\mu \\
    \texttt{ncc[thetabar, thetabar]\,ncc[theta, omega] +}& \leftrightarrow
      \bar \theta \!\cdot\! \bar \theta\ \theta \!\cdot\! \omega \\
    \texttt{ncc[theta, theta] \,ncc[thetabar, rhobar] +} &
    \leftrightarrow \theta \!\cdot\! \theta\ \bar \theta \!\cdot\! \bar \rho \\
    \texttt{ncc[theta, theta] \,ncc[thetabar, thetabar]\,d}
    &\leftrightarrow \theta \!\cdot\! \theta\ \bar \theta \!\cdot\! \bar \theta
      d \ .
   \end{array}
\esp\ee

The number of
degrees of freedom of a generic superfield is
too large to match those of the supermultiplets representing the
$N=1$ supersymmetric algebra. Two special cases, with fewer degrees of freedom,
are in general enough to build most of the phenomenologically relevant
supersymmetric theories.
These so-called chiral and vector superfields will be discussed in the next sections.

\subsection{Chiral superfields} \label{sec:Csuperfields}
Left and right-handed chiral superfields are superfields $\Phi_L$ and
$\Phi_R$ that satisfy the constraints
\be
  \bar D_\alphadot \Phi_L (x,\theta,\bar\theta) = 0 
  \quad \text{and} \quad
  D_\alpha \Phi_R (x,\theta,\bar\theta) = 0 \ ,
\ee 
where the superderivatives have been introduced in Eq.\
\eqref{eq:superch-and-der}.  Since the superderivatives anticommute with the supercharges, the constraints are preserved by supersymmetry transformations. The most general solutions to the constraints can be
written as
\be\bsp
  \Phi_L(y,\theta) = &\ \phi(y) + \sqrt{2} \theta \cdot \psi(y) - \theta \cdot
    \theta F(y) \ , \\
  \Phi_R(y^\dagger,\bar\theta) = &\ \phi(y^\dagger) + \sqrt{2} \bar\theta \cdot
    \bar\psi(y^\dagger) - \bar\theta \cdot \bar\theta F(y^\dagger) \ ,
\esp\label{eq:chiralSF} \ee
where we have introduced the variable\footnote{The $\sqrt{2}$ factors and the minus signs are purely
conventional.} $y^\mu = x^\mu - i\theta \sigma^\mu \bar
\theta$ and its hermitian conjugate $(y^\mu)^\dag = x^\mu + i\theta \sigma^\mu \bar
\theta$. Chiral superfields are appropriate to describe the $N=1$ matter
supermultiplets containing one
fermion and one scalar field. Indeed, the physical degrees of freedom of $\Phi_L$ and
$\Phi_R$ consist of a single complex scalar field $\phi$ and a single
two-component fermion $\psi$, the chirality of the
Weyl fermion providing the name for this kind of superfield. The additional scalar field
$F$ has a mass dimension $[F]=2$ and does hence not correspond to a physical degree of freedom,
but it is necessary to restore the
equality between the numbers of fermionic and bosonic degrees of freedom off-shell.
Furthermore, these so-called $F$-terms can be used to break supersymmetry
spontaneously if they develop a vacuum expectation value. 
As we will discuss in Section~\ref{sec:autolag}, these fields are non-propagating
and can be eliminated through their equations of motion. 

Left and right-handed chiral superfields
can be declared in \feynrules\ following in the \feynrules\ model file in a way similar to the ordinary fields,
\eg,
\begin{verbatim}
M$Superfields = {
  CSF[1] == {
    ClassName -> PHI,
    Chirality -> Left,
    Weyl      -> psi,
    Scalar    -> z,
    Auxiliary -> FF},
  
  CSF[2] == {
    ClassName -> OMEGA,
    Chirality -> Right,
    Weyl      -> xibar,
    Scalar    -> zz} 
}  .
\end{verbatim}
These \mathematica\ commands declare two chiral superfields,
 the label of the particle class {\tt CSF} referring to chiral superfields.
We consider the example of both a left and right-handed chiral superfield called {\tt PHI} and
 {\tt OMEGA}. The fermionic degrees of freedom of the superfields are {\tt psi} and {\tt xibar}, 
 while the scalar degrees of freedom are denoted by {\tt z} and {\tt zz}. In
addition to all the options available to all particle classes and briefly reviewed in Section~\ref{sec:FR}, such as {\tt
QuantumNumbers} and {\tt Indices}, the user must set the {\tt
Chirality} option to {\tt Left} or
{\tt Right}, related to the chirality of the superfield.
Moreover, 
it is mandatory to link a chiral superfield to its fermionic and scalar components
through the options {\tt Weyl} and {\tt Scalar}, respectively. Note that
the component fields have to be declared independently in the \feynrules\ model
file (see Section~\ref{sec:FR} for a brief review or
Refs.~\cite{Christensen:2008py,Butterworth:2010ym} for a detailed description). 
We only emphasize here that the options for the component field classes, 
like {\tt Indices, QuantumNumber}, \emph{etc.}, must be identical to the corresponding options
of the superfield. The only exception to this rule are the $F$-terms, where the user can either point 
to an already declared complex scalar field via the {\tt Auxiliary} options (as for {\tt CSF[1]} 
in the example above, where the auxiliary field is denoted {\tt FF}), or he
leaves this option 
unspecified and \feynrules\ creates internally a symbol for the auxiliary field (as for {\tt CSF[2]} 
in the example above).
All the allowed options for the declaration of a chiral superfield are
summarized in Table \ref{tab:CSF}.

\renewcommand{\arraystretch}{1.5}
\begin{table}
\bgfbalign
\multicolumn{2}{c}{\textbf{Table~\ref{tab:CSF}: Chiral superfield class options}}\\
{\tt ClassName} & Defines the symbol by which a class is represented.\\
{\tt Chirality} & Defines the chirality, {\tt Left} or {\tt Right},
  of the chiral superfield. \\
{\tt Weyl} & Contains the symbol of the fermionic component of the
  chiral superfield.\\
{\tt Scalar} & Contains the symbol of the scalar component of the
chiral superfield.\\
\multicolumn{2}{l}{\textbf{Optional attributes}}\\
{\tt Auxiliary} &  Contains the symbol of the auxiliary component of
the chiral superfield. If absent, \feynrules\ generates the $F$-term
automatically.\\
{\tt Indices} & The list of indices, different from Lorentz and spin
  indices, carried by the superfield and all its component fields.\\
{\tt QuantumNumbers} & A replacement rule list, containing the $U(1)$ quantum numbers
carried by the class.\\
\egfbalign
\textcolor{white}{\caption{\label{tab:CSF}}}
\end{table}
\renewcommand{\arraystretch}{1}

\subsection{Vector superfields in the Wess-Zumino gauge}\label{sec:Vsuperfields}
Besides matter fields, gauge theories contain also real vector fields describing the gauge bosons. 
Since the chiral superfields defined in the previous section do not contain any vector degree of freedom, they cannot be sufficient to describe supersymmetric gauge theories. We therefore 
introduce in this section the so-called 
vector (or gauge) supermultiplets, \ie, the representations of the $N=1$ SUSY
algebra containing one massless gauge boson together with the corresponding
fermionic degree of freedom. These multiplets can be described by vector
superfields, defined by the reality condition,
\be
  V = V^\dagger \ .
\label{eq:reality}\ee
The constraint \eqref{eq:reality} on its own is not enough to reduce the number of degrees of
freedom of the generic superfield of Eq.\ \eqref{eq:genSFa} to the required
number and leads to a proliferation of unphysical fields
that can be eliminated by a suitable gauge choice.
A convenient choice is the so-called Wess and Zumino
gauge, in which a vector superfield is expressed as
\be
  V_{W.Z.} = \theta \sigma^\mu \bar \theta v_\mu + 
    i \theta \cdot \theta\ \bar \theta \cdot \bar \lambda - 
    i \bar \theta \cdot \bar \theta\ \theta \cdot \lambda + 
    \frac12 \theta \cdot \theta\ \bar \theta \cdot \bar \theta D \ .
\ee
The component fields of a vector superfield are hence a real vector field
$v_\mu$ and a Majorana fermion $(\lambda_\alpha,\bar\lambda^\alphadot)$. Similar
to the case of the chiral superfield, the multiplet also contains a
non-propagating auxiliary scalar field $D$ with mass  dimension
$[D]=2$ that is necessary
to restore the equality between the numbers of fermionic and bosonic
degrees of freedom off-shell. These so-called $D$-terms can again be eliminated  through 
their equations of motion.

 Vector superfields in the Wess and Zumino gauge can be implemented into \feynrules\ in the same way as chiral superfield. Let us illustrate this by an example,
\begin{verbatim}
M$Superfields = {
  VSF[1] == {
    ClassName        -> PHIV,
    GaugeBoson       -> X,
    Gaugino          -> lambda}
} . \end{verbatim}
In this example, {\tt PHIV} denotes a vector superfield, whose vector component is
denoted by {\tt X} and the fermionic component by {\tt lambda}. The label of the particle
class {\tt VSF} refers to its vector superfield nature.
In addition to the usual options for particle classes, the user must associate
to a vector superfield its bosonic and fermionic components
by setting the options {\tt GaugeBoson} and {\tt Gaugino}. Note that the corresponding vector field and the Weyl fermion must be declared 
independently in the \feynrules\ model file. As in the case of chiral superfields, the 
$D$-terms can either be defined explicitly via the {\tt Auxiliary} option,
or this option can be omitted and \feynrules\ will declare the $D$-terms
internally. All the allowed options for the declaration of a vector superfield are
summarized in Table \ref{tab:VSF}.
In addition, each vector superfield can be linked to a gauge group
through the option {\tt
Superfield} which has been added to the gauge group class in a similar way as the we assigned the gluon field {\tt G} to the QCD gauge group {\tt SU3C} in Section~\ref{sec:FR}. As an example, we
could associate the superfield {\tt PHIV} above to an abelian gauge group called
{\tt U1X} by declaring the gauge group as
\begin{verbatim}
  U1X == {
    Abelian          -> True,
    CouplingConstant -> gX, 
    Superfield       -> PHIV}  .
\end{verbatim}

\renewcommand{\arraystretch}{1.5}
\begin{table}
\bgfbalign
\multicolumn{2}{c}{\textbf{Table~\ref{tab:VSF}: Vector superfield class options}}\\
{\tt ClassName} & Defines the symbol by which a class is represented.\\
{\tt GaugeBoson} & Contains the symbol of the vector field
associated to the vector superfield.\\
{\tt Gaugino} & Contains the symbol of the gaugino component of the
vector superfield.\\
\multicolumn{2}{l}{\textbf{Optional attributes}}\\
{\tt Auxiliary} &  Contains the symbol of the auxiliary component of
the superfield. If absent, \feynrules\ generates the $D$-term
automatically.\\
{\tt Indices} & The list of indices, different from Lorentz and spin
  indices, carried by the superfield and all its component fields.\\
\multicolumn{2}{l}{\textbf{New gauge group options}}\\
{\tt Superfield} & This option points to the {\tt ClassName} of the vector superfield
  associated to the gauge group. If the {\tt Superfield} option is present, the
  option {\tt GaugeBoson} becomes optional. If both options are there, they must
  be consistent.\\
\egfbalign
\textcolor{white}{\caption{\label{tab:VSF}}}
\end{table}
\renewcommand{\arraystretch}{1}

\section{Simplification of superspace expressions}
\label{sec:simpli}

\subsection{Simplification of expressions involving Grassmann numbers}
\label{sec:sim1}
In the previous sections we have discussed the implementation of superfields into \feynrules. 
The user can either use
the predefined classes {\tt CSF} and {\tt VSF} for chiral and vector superfields, 
or implement superfields directly in terms of the component fields, as in
the example of
Eq.\ \eqref{eq:genSF}. In that case, however, the required synthax for the
implementation can be complex, and as a consequence the \mathematica\ output
could be difficult to read. Similarly, many calculations such as
those involving the supersymmetric generators or the superderivatives could
lead to rather long expressions after introducing the second rank
antisymmetric tensors and the {\tt nc} environment, which are necessary in
order to have the fermions ordered correctly and all the spin indices carried
by fields lowered, 
as is the convention in the superspace module of \feynrules. To
bypass this issue, it might be useful to form
dot products of spinors, and more specifically
those involving the Grassmann variables $\theta$ and $\bar\theta$. This can
be achieved with the help of the {\tt ToGrassmannBasis} function, which proceeds in
two steps. First, dot products are formed and simplified using relations among
the Grassmann variables, such as
\be\label{eq:canonical_form}
  \theta^\alpha \theta^\beta = -\frac12\theta\cdot\theta \e^{\alpha \beta}\ , \ 
  \bar\theta^\alphadot\bar\theta^\betadot = \frac12\bar\theta\cdot\bar\theta
    \e^{\alphadot \betadot} \ , \
  \theta^\alpha \bar\theta^\alphadot = \frac12\theta \sigma^\mu
    \bar\theta\bar\sigma_\mu{}^{\alphadot \alpha} \ , \ \ldots 
\ee 
After this step, any function of the component fields is written in terms of a
restricted set of scalar products involving Grassmann variables and Pauli
matrices, forming hence a basis in which any superfield expression can be
expanded. 

Even though this procedure in principle solves the problem of simplifying superfield expressions,
we have to deal at this stage with a purely technical issue. Expressing
everything in terms of
the basic objects results in \mathematica\ expressions that are equal up to the names of contracted indices, \emph{e.g.},
\be \texttt{
  Dot[theta[al], theta[al]] -  Dot[theta[be], theta[be]] } \ . 
\label{eq:doubledot}\ee
Although this difference is manifestly zero, the two terms represent 
different patterns in
\mathematica, and hence the cancellation does not take place. In general, such a
situation can occur relatively often with the {\tt ToGrassmann\-Basis} function, and therefore the internal index naming
scheme is optimized by the {\tt ToGrassmannBasis} function after all the dot products are formed. Similar terms are
collected and summed, and hence the readability of the final
results is highly improved. The simplification function presented above can
be called in \feynrules\ through the command
\be 
 \texttt{ToGrassmannBasis[expression] } \ , 
\nn\ee
where {\tt expression} is any function of the component fields. The
{\tt ToGrass\-mann\-Basis} function then expresses {\tt expression} in the basis 
defined by the objects in Eq.~\eqref{eq:canonical_form}. 

As an example, the function {\tt ToGrassmannBasis} could significantly help to
improve the readability of the most general series in the $\theta$ and
$\bar\theta$ variables presented in Eq.\ \eqref{eq:genSF}. Using the
\mathematica\ variable {\tt Phi} defined in Eq.\ \eqref{eq:genSF}, the command
\be
   \texttt{ToGrassmannBasis[Phi]} 
\nn\ee 
allows the user to obtain an expression much closer to the original form
of Eq.\ \eqref{eq:genSFa},  
\be\bsp
 & \texttt{z + theta[sp].xi[sp] + zetabar[spd].thetabar[spd]\! +}\\
 &\ \texttt{
    f*theta[sp].theta[sp]+g*thetabar[spd].thetabar[spd]\! +}\\
 &\ \texttt{
    theta[sp].thetabar[spd]*si[mu,sp,spd]*V[mu]\! +}\\
 &\ \texttt{
    rhobar[spd].thetabar[spd]*theta[sp].theta[sp]\! +}\\
 &\ \texttt{
    theta[sp].omega[sp]*thetabar[spd].thetabar[spd]\! +}\\
 &\ \texttt{
    d*theta[sp].theta[sp]*thetabar[spd].thetabar[spd]} \ .
\esp \label{eq:simplsup} \ee

It is important to note that the {\tt ToGrassmannBasis} function can also work
on spinorial or tensorial expressions, \ie, expressions with uncontracted spin
(or vectorial) indices. After an application of the simplification
module, the (upper or lower) free index could be attached to a single fermion,
or to a chain containing one fermion and a given number of Pauli matrices,
\be
  \chi^\alpha \ , \quad 
  \chi^\alpha \sigma^\mu{}_{\alpha\alphadot} \leftrightarrow 
    \big(\chi \sigma^\mu\big)_\alphadot \ , \quad
  \sigma^\mu{}_{\alpha\alphadot}
    \bar\sigma^{\nu\alphadot\beta} \chi_\beta \leftrightarrow 
    \big(\sigma^\mu\bar\sigma^\nu\chi\big)_\alpha \ , \quad
  \ldots \ ,
\label{eq:chains}\ee
where, in the examples above, $\chi$ denotes a generic left-handed Weyl fermion.
The {\tt ToGrassmannBasis} function has been implemented so that those chains
are formed and stored in the {\tt TensDot2} environment following the
pattern
\be 
  \texttt{TensDot2[ chain ][pos, chir, name]} \ , 
\nn\ee
where {\tt chain} is a sequence of one Weyl fermion and possibly one or several
Pauli matrices, {\tt pos} is the {\tt up} or {\tt down} position of the free spin
index, {\tt chir} its chirality, \ie, its dotted or undotted nature, and {\tt
name} its name. The three examples of Eq.\ \eqref{eq:chains} could be
implemented as
\be \bsp
  \chi^\alpha \leftrightarrow &\ 
     \texttt{ToGrassmannBasis[ nc[chi[b]] * Ueps[a,b] ]} \ , \\
  \big(\chi \sigma^\mu\big)_\alphadot \leftrightarrow &\ 
     \texttt{ToGrassmannBasis[ nc[chi[b]] * si[mu,a,ad] *}\\
     &\qquad \texttt{Ueps[a,b] ]} \ , \\
  \big(\sigma^\mu \bar\sigma^\nu\chi\big)_\alpha
    \leftrightarrow &\
     \texttt{ToGrassmannBasis[ nc[chi[b]] * si[mu,a,ad] *}\\
     &\qquad \texttt{sibar[nu,ad,b] ]} \ .
\esp\nn\ee
We recall that the {\tt nc} environment is mandatory as soon as we are
dealing with fermions and that by convention all spin indices carried by
fermionic fields are considered to be lowered. 
We then obtain
\be \bsp
  &\ \texttt{nc[ TensDot2[chi[a]][up,Left,a] ] } \ ,\\
  &\ \texttt{nc[ TensDot2[chi[a], si[mu,a,ad]][down,Right,ad] ]} \ ,\\
  &\ \texttt{nc[ TensDot2[si[mu,a,ad], sibar[nu,ad,b],}\\
  &\qquad \texttt{chi[b]][down,Left,a] ]} \ .
\esp \nn \ee

To conclude this section, let us comment on the index optimization routine used by the {\tt 
ToGrassmannBasis} function, because there might be times where the user wants to handle the 
optimization of the
index naming scheme without forming scalar products involving Grassmann
variables, \ie, without calling the {\tt ToGrassmannBasis} function. The standalone version of the
 index optimization that 
consistently renames the indices of an expression is called as 
\be
  \texttt{OptimizeIndex[expression, list] } \ ,
\nn\ee
where {\tt list} is an optional list of variables carrying indices to be included into the index renaming   and that
are neither a field nor a parameter included in the global variable of
\feynrules\ {\tt M\$Parameters}.
As an example, the application of the optimization of the index naming scheme on
the sum of scalar products of Eq.\ \eqref{eq:doubledot} can be performed with
the command
\be \bsp
  &\ \texttt{OptimizeIndex[ Dot[theta[al],theta[al]] -   } \\
  &\ \qquad \texttt{ Dot[theta[be],theta[be]]] } \ , 
\esp \nn \ee
which simplifies to zero. Equal terms have now been consistently
subtracted and are not repeated anymore.

\subsection{Simplified input format for superfield expressions} \label{sec:sim2}
In Section~\ref{sec:superspace} we introduced a canonical form in which every superspace function can be expressed. This canonical form is defined by the two simple rules
\begin{enumerate}
\item all spin indices carried by fermion fields are considered lowered,
\item the ordering of the fermions is implemented via the {\tt nc} environment.
\end{enumerate} 
As we already discussed, this canonical form usually requires the explicit use
of the $\epsilon$-tensors, and for this reason, we have introduced in the previous
section the {\tt ToGrassmannBasis} function
which reduces any superfield expression to the basis of Grassmann variable monomials defined by 
Eq.\ \eqref{eq:canonical_form}. Conversely, this basis can also be used to input
superspace
expressions in an easier way. The rules for this input format are discussed in the rest of this section.

First, the dot products of spinors, connected or not with
the help of Pauli matrices, are always written as
\be
  \texttt{ ferm1[sp1].ferm2[sp2] chain[sp1,sp2] } ,
\nn\ee
where the symbols {\tt ferm1} and {\tt ferm2} denote the two fermions and 
{\tt chain} contains a series of Pauli matrices linking the two spin
indices {\tt sp1} and {\tt sp2}. As a simple example, a possible connected
product of two left-handed spinors $\chi$ and $\psi$ would be
\be
  \chi \sigma^\mu \bar\sigma^\nu \psi \leftrightarrow 
  \texttt{chi[sp1].psi[sp2] si[mu,sp1,spd] sibar[nu,spd,sp2]} \ ,
\nn\ee
the symbol {\tt chain} being here equal to {\tt si[mu,sp1,spd]
sibar[nu,spd,sp2]}. In the cases where there is no chain of
Pauli matrices present in the expression, the spin indices carried by the
fermions must be equal, the dot product being hence a regular scalar product of
fermions, as those introduced in Eq.\ \eqref{eq:scalprod}. A few additional
examples can be found among the different terms of Eq.\ \eqref{eq:simplsup}.
Second, any fermionic expression carrying a free spin index must use both the
{\tt nc} environment as well as the {\tt TensDot2} structure, following the
synthax explained in Section \ref{sec:sim1}.

Even though this format allows to input superspace expressions in an easier way, most of the 
functionalities of the superspace module require the input expressions to be given in the canonical 
form of Section~\ref{sec:superspace}. In the next section we therefore introduce a function that allows
to convert an expression written in terms of the basis objects of Eq.~\eqref{eq:canonical_form} to its canonical form.

\subsection{Reverting simplifications: back to the {\tt nc} environment}
One drawback of the optimization routines is that several of the functions
included in the superspace module of \feynrules, such as the {\tt QSUSY} or {\tt
DSUSY} routines presented in
Section \ref{sec:superspace}, require expressions including the {\tt nc}
environment, second rank antisymmetric tensors, lower spin indices for fermions
and not any explicit scalar products. A simple solution is provided by the
{\tt Tonc} function. Indeed, if one needs to perform
additional operations on simplified expression within the superspace, it is
recommended to first use the {\tt Tonc} function, which allows to transform
simplified expressions back to their original form in terms of the {\tt nc}
environment and the epsilon tensors.
Hence, {\tt Tonc[Dot[theta[a],theta[a]]]} would lead to
\be \texttt{
  nc[theta[a],theta[sp\$1]] Ueps[sp\$1,a]
} \ ,
\nn\ee
where the second spin index has been automatically generated by \feynrules.
Sometimes, it might be useful to call the {\tt OptimizeIndex} function right
after the use of the {\tt Tonc} module in order to get shorter and simpler
expressions.

Note that the {\tt Tonc} function allows us at the same time to enter expressions directly with the use of the
simplified synthax presented in Sections \ref{sec:sim1} and \ref{sec:sim2}, and to
convert them to their canonical form.

\renewcommand{\arraystretch}{1.5}
\begin{table}
\bgfbalign
\multicolumn{2}{c}{\textbf{Table~\ref{tab:simpli}: Simplification functions}}\\
{\tt ToGrassmannBasis[exp]} &\\
 & This function allows to express any function {\tt exp}
of the component fields in the basis of Grassmann monomial defined by Eq.~\eqref{eq:canonical_form}. \\
{\tt OptimizeIndex[exp,list]} &\\
 & This function allows for the optimization of the index naming scheme used in
  the expression {\tt exp}. The optional argument {\tt list} is a list of
  variables carrying indices, which are neither a field nor included in
  {\tt M\$Parameters}.\\
{\tt Tonc[exp]} & This function transforms simplified expressions back to their
  original form, with lower spin indices, epsilon tensors, {\it etc}...\\
\multicolumn{2}{l}{\textbf{New environments}}\\
{\tt TensDot2[chain][pos,chir,name]} &\\
& This environment contains a sequence, labelled by {\tt chain}, of one Weyl
fermion and possibly one or several Pauli matrices. The symbols {\tt pos}, {\tt
chir} and {\tt name} are the {\tt up} or {\tt down} position, the chirality and
the name of the free index.\\
\egfbalign
\textcolor{white}{\caption{\label{tab:simpli}}}
\end{table}
\renewcommand{\arraystretch}{1}

\section{Manipulating superfield expressions}
\label{sec:SF2Parts}

\subsection{From superfields to Lagrangians}
The main advantage of writing down supersymmetric Lagrangians in terms of
(chiral and vector) superfields rather than in terms of the component fields is
the size of the corresponding expressions. This is illustrated in the
following simple example. Let us consider a left-handed chiral superfield
$\Phi$ whose scalar, fermionic and auxiliary
components are denoted by $\phi$, $\psi$ and $F$. Performing the series expansion of $\Phi^\dag\Phi$ in terms of the
Grassmann variables, it can be shown that the coefficient with the
highest power in $\theta$ and $\bar\theta$, \ie, the coefficient of the $\theta\cdot\theta
\bar\theta\cdot\bar\theta$ term, is SUSY invariant and hence a good candidate for a Lagrangian density
describing the free component fields. Indeed, we get 
\be\bsp
  \lag = \Phi^\dag \Phi_{|_{\theta\cdot\theta \bar\theta\cdot\bar\theta}} =&\ 
    -\frac14 \Big(\phi^\dag \Box \phi + \Box \phi^\dag \phi - 2
     \del_\mu \phi^\dag \del^\mu \phi \Big) +  F^\dag F  \\
   &\ + \frac{i}{2} \Big(\psi \sigma^\mu \del_\mu \bar \psi - \del_\mu \psi \sigma^\mu
      \bar \psi\Big)\ .
\esp\label{eq:freelag}\ee

The expansion of a superfield polynomial into a series in the Grassmann variables 
can be automatically performed in \feynrules\ via the {\tt SF2\-Components}
function. Since this function allows in the same way to transform superfield 
expressions into four-dimensional Lagrangians, it is one of the most important
functions available and at the heart of the superspace module of \feynrules.
For any given function of chiral and vector superfields denoted {\tt expression},
the correct syntax to use is simply
\be
  \texttt{SF2Components [ expression ] } \ .
\nn\ee
The {\tt SF2Components} function expands all the superfields appearing
in {\tt ex\-pres\-sion} in terms of their component fields and the
$x^\mu$ spacetime coordinates (rather than the $y^\mu$ variable related to 
chiral superfields). In a second step, scalar products
of Grassmann variables are simplified and the expression is reduced to the basis defined by Eq.~\eqref{eq:canonical_form} using the {\tt ToGrassmannBasis} function. 
During this procedure representation
matrices of the Lie algebra of the gauge groups are introduced and simplified, using, \eg, the commutation relations 
between the generators.
The output of the {\tt SF2Components} function consists in a list of two elements,
\be
  \texttt{ \{ Full series , List of the nine coefficients \} } \ .
\label{eq:SF2CompOutput}\ee
The first element of this list, labelled here {\tt Full series}, is the full
series expansion in the Grassmann
variables which could also have been directly obtained with the {\tt GrassmannExpand}
function,
\be \texttt{
  GrassmannExpand[ expression ] } \ .
\nn \ee
The second element
of the list of Eq.\ \eqref{eq:SF2CompOutput} is a list containing the
nine coefficients of the series. The
first part of this last list is the scalar piece of the series, \ie\ the terms
independent of the $\theta$ and $\bar\theta$ variables, while the other
elements are the coefficients of the $\theta_\alpha$, $\bar\theta_\alphadot$,
$\theta \sigma^\mu\bar\theta$, $\theta\cdot\theta$, $\bar\theta\cdot\bar\theta$,
$\theta\cdot\theta \bar\theta_\alphadot$, $\bar\theta\cdot\bar\theta
\theta_\alpha$ and $\theta\cdot\theta \bar\theta\cdot\bar\theta$ terms,
following this ordering. Each of them can also be obtained using the functions,
\be\bsp
&\  \texttt{ScalarComponent [ expression ] } \ , \\
&\  \texttt{ThetaComponent [ expression ] } \ , \\
&\  \texttt{ThetabarComponent [ expression ] } \ , \\
&\  \texttt{ThetaThetabarComponent [ expression ] } \ , \\
&\  \texttt{Theta2Component [ expression ] } \ , \\
&\  \texttt{Thetabar2Component [ expression ] } \ , \\
&\  \texttt{Theta2ThetabarComponent [ expression ] } \ , \\
&\  \texttt{Thetabar2ThetaComponent [ expression ] } \ , \\
&\  \texttt{Theta2Thetabar2Component [ expression ] } \ .
\esp\label{eq:shortcut}\ee
The results obtained by each of these functions are thus all independent of the Grassmann
variables. It is important to note that each
of these functions calls the {\tt SF2Components} module, the complete series being
thus recalculated at each function call. Hence, if several of the coefficients have to be
computed, it is much faster to store and re-use the results of the {\tt
SF2Components} function than to call all the {\tt XXXXComponent} functions
individually. Furthermore, these functions can also be used on expressions in
terms of the component fields. In this case, we recall that the use of the {\tt Tonc}
environment is mandatory. All the available functions
to reexpress superfield expressions in terms of their component fields are
summarized in Table \ref{tab:SF2Part}.

\renewcommand{\arraystretch}{1.5}
\begin{table}
\bgfbalignstretch
\multicolumn{2}{c}{\textbf{Table~\ref{tab:SF2Part}: From superfields to
particles}}\\
{\tt SF2Components[exp]} 
  &Expands the superfield expression {\tt
    exp} in terms of its component fields and simplifies the products of Grassmann
    variables. The result is a list of two
    elements, the complete series and a new list with all the individual
    coefficients.\\
\multicolumn{2}{l}{\textbf{Shortcuts to the individual component fields}}\\
The full series. & {\tt GrassmannExpand[exp]} \\
The scalar term. & {\tt ScalarComponent[exp]}\\
The $\theta$ term. & \texttt{ThetaComponent[exp]} \\
The $\bar\theta$ term. & \texttt{ThetabarComponent[exp]} \\
The $\theta\sigma\bar\theta$ term.&  \texttt{ThetaThetabarComponent[exp]}\\
The $\theta^2$ term.& \texttt{Theta2Component[exp]}\\
The $\bar\theta^2$ term.& \texttt{Thetabar2Component[exp]} \\
The $\theta^2\bar\theta$ term. & \texttt{Theta2ThetabarComponent[exp]}\\
The $\bar\theta^2\theta$ term.&\texttt{Thetabar2ThetaComponent[exp]}\\
The $\theta^2\bar\theta^2$ term. & \texttt{Theta2Thetabar2Component[exp]}
\egfbalignstretch
\textcolor{white}{\caption{\label{tab:SF2Part}}}
\end{table}
\renewcommand{\arraystretch}{1}

As an example, let us derive the Lagrangian of Eq.\ \eqref{eq:freelag} with
\feynrules, starting from the superfield expression, recalling that the
superfield {\tt PHI} has been defined in Section \ref{sec:Csuperfields}.
This can be done via the \mathematica\ command
\be
   \texttt{Theta2Thetabar2Component[PHIbar PHI]} \ ,
\nn \ee 
and one obtains indeed the correct free Lagrangian for the component fields, 
\be\bsp
  &\ \texttt{(del[z, mu]*del[zbar, mu])/2 - }\\
  & \qquad\texttt{(zbar*del[del[z, mu], mu])/4 -}\\
  & \qquad\texttt{(z*del[del[zbar, mu], mu])/4 + FF*FFbar -}\\ 
  & \qquad\texttt{(I/2)*del[xi[sp],mu].xibar[spd] * 
    si[mu,sp,spd] +}\\ 
  & \qquad\texttt{(I/2)*xi[sp].del[xibar[spd],mu] * 
    si[mu,sp,spd]
} \ , 
\esp\label{eq:chirallagout}\ee
where {\tt del} stands for the spacetime derivative.

\subsection{Supersymmetric transformation laws}
\label{sec:SUSY-laws}
The supersymmetric transformation laws of
a superfield can be obtained by using the explicit representation of the
supercharges given in Eq.\ \eqref{eq:superch-and-der}. Hence, considering an
infinitesimal supersymmetric transformation of a (spinorial) parameter
$(\epsilon_\alpha, \bar\epsilon^\alphadot)$, a superfield transforms as
\be
  \Phi \to \Phi + \delta_\epsilon \Phi = \Phi + i \big(\epsilon\cdot Q + \bar
    Q\cdot\bar\epsilon\big) \Phi \ .
\label{eq:SUSYtransf}\ee
After expanding the superfield in terms of its component fields and using Eq.\
\eqref{eq:superch-and-der}, we can immediately read off the transformations of
the component fields.
The operator $\delta_\epsilon$ is implemented into the superfield module via
the {\tt DeltaSUSY} function,
\be
  \texttt{ DeltaSUSY [ expression , epsilon ] } \ ,
\nn \ee
 where {\tt expression} can be any function of superfields and/or component
fields and {\tt epsilon} refers to the supersymmetric transformation
parameter, without any spin index. There are ten such parameters
predefined in the superfield module, labelled by {\tt eps\it{x}} with
{\tt\it{x}} being an integer between zero and nine, which the user can use at
his convenience. This parameter being a Majorana fermion,
it is enough to only provide the associated two-component left-handed
spinor\footnote{The objects {\tt eps\it{x}} are in fact implemented into \feynrules\ as 
left-handed Weyl fermions. The corresponding right-handed objects are also available under the name {\tt eps\it{x}bar}.}. The output of the {\tt DeltaSUSY} module is the full series expansion in
the Grassmann variables. Finally, the functions given in Eq.\ \eqref{eq:shortcut} allow for the
identification of the variations of the various component fields.

Let us illustrate the use of this function on the explicit example of chiral superfields, 
and let us start with the superfield {\tt PHI} introduced in Section
\ref{sec:Csuperfields}. We then calculate its variation under a supersymmetric
transformation of parameter $\epsilon_1$, 
\be\bsp
  \texttt{DeltaPHI = DeltaSUSY[ PHI , eps1 ]} \ , 
\esp\nn\ee
where we have used the supersymmetric transformation parameter {\tt eps1}. 
The scalar, $\theta_\alpha$ and $\theta\cdot\theta$ coefficients of {\tt
DeltaPHI} are finally identified with the variations of the scalar component
$\delta_{\epsilon_1} z$, the fermionic component $\delta_{\epsilon_1} \psi$
and of the auxiliary component
$\delta_{\epsilon_1} F$ of the superfield. The individual components can be
easily extracted using the shortcuts of  Eq.\ \eqref{eq:shortcut}, and
one recovers the well-known textbook expressions,
\be\bsp
  \delta_{\epsilon_1} z =&\ \sqrt{2} \epsilon_1 \cdot \psi \ ,\\
  \delta_{\epsilon_1} \psi =&\ -\sqrt{2} \epsilon_1 F -i \sqrt{2} \sigma^\mu \bar
    \epsilon_1 \del_\mu z \ , \\
 \delta_{\epsilon_1} F =&\ -i \sqrt{2} \del_\mu \psi \sigma^\mu \bar \epsilon_1 \ , 
\esp\ee
where $z$, $\psi$ and $F$ are the scalar, fermionic and auxiliary component of
the considered superfield.

\section{Implementing supersymmetric Lagrangians into \feynrules}
\label{sec:lagrangians}

\subsection{Supersymmetric Lagrangians}\label{sec:laggen}
In this section we describe how to generate in an automated way (parts of) the
Lagrangian describing the interactions between the different component fields of
chiral and vector supermultiplets. The first part of this section is devoted to a
brief review on how to construct supersymmetric Lagrangians.
The most general supersymmetry-conserving
Lagrangian describing the interactions between the various multiplets can be
written as a sum of three pieces,
\be
  \lag = \lag_{\rm chiral} + \lag_{\rm Yang-Mills} + \lag_{\rm superW}  \ ,
\label{eq:SUSYlag}\ee
where $\lag_{\rm chiral}$ and $\lag_{\rm Yang-Mills}$ contain the kinetic terms
as well as the gauge interactions of the different particles, and $\lag_{\rm superW}$ is the superpotential describing the interactions between the different chiral supermultiplets.
Note that, since supersymmetric particles have not yet been
observed, supersymmetry must be broken at low energies, which renders the
superpartners heavy in comparison to their Standard Model counterparts. The
corresponding Lagrangian does not involve superfields, but only some of the
component fields, and so we exclude it from the present discussion and
only concentrate on Lagrangians where supersymmetry is unbroken.

The first Lagrangian in Eq.\ \eqref{eq:SUSYlag}, $\lag_{\rm chiral}$,
contains the kinetic terms as well as the gauge interactions of the chiral superfields. 
It is completely fixed by gauge invariance and reads
\be
  \lag_{\rm chiral} = \bigg[ \Phi_i^\dag \, e^{-2 g_j V^j} \,
    \Phi^i \bigg]_{\theta\cdot\theta \bar\theta\cdot\bar\theta}\ ,
\label{eq:chiralSFLag}\ee
where a sum over all chiral superfields $\Phi^i$ and vector superfields $V^j =
V^{ja}\,T_a$ is understood. Furthermore, $T_a$ denote the representation
matrices of the supermultiplet $\Phi^i$ and $g_j$ is the associated gauge
coupling constant.
It can be shown that the
$\theta\!\cdot\!\theta \bar\theta\!\cdot\!\bar\theta$ coefficient of the
expansion in the Grassmann variables of the expression inside the
squared brackets is invariant under supersymmetry transformations
and reproduces for gauge singlet superfields the Lagrangian~\eqref{eq:freelag} 
for a free chiral supermultiplet. This expression is hence a good candidate for
a Lagrangian density. In general, if the fields carry some gauge charges, 
the expansion of Eq.~\eqref{eq:chiralSFLag} yields
\be\bsp
  \lag_{\rm chiral} &\ = 
      D_\mu \phi_i^\dag D^\mu\phi^i + F_i^\dag F^i
    - \frac{i}{2} \big(D_\mu \bar \psi_i \bar \sigma^\mu \psi^i - 
        \bar \psi_i\bar \sigma^\mu  D_\mu \psi^i \big) \\
  &  + 
      i\sqrt{2}g_j  \bar \lambda^{ja} \cdot \bar \psi_i T_a \phi^i  
    - i\sqrt{2}g_j \phi_i^\dag T_a  \psi^i \cdot  \lambda^{ja} - g_j D^{ja}
      \phi_i^\dag T_a \phi^i \  ,
\esp \ee
where $(\phi^i, \psi^i, F^i)$ and $(V^j_\mu, \lambda^j, D^j)$ denote the
component fields of the chiral and vector supermultiplets $\Phi^i$ and $V^j$,
and $D_\mu$ is the covariant derivative
\be
  D_\mu = \del_\mu - i g_j V^{ja}_\mu T_a\ .
\ee

The second Lagrangian of Eq.\ \eqref{eq:SUSYlag}, $\lag_{\rm Yang-Mills}$,
contains the kinetic terms and self-interactions of the vector superfields. It
consists in a sum over all the vector supermultiplets of the theory, and can be
written, for one specific vector supermultiplet $V$,
\be
\lag_{{\rm Yang-Mills}, V} = 
   \frac{1}{16 g^2} \Big[W^\alpha{}_a
     W_\alpha{}^a \Big]_{\theta\cdot \theta } +
   \frac{1}{16 g^2} \Big[\overline W_\alphadot{}^a
     \overline W^\alphadot{}_a \Big]_{ \bar \theta\cdot\bar\theta} \ .
\label{eq:LVec}\ee
Similarly to $\lag_{{\rm chiral}}$, the squared brackets indicate that we only
take the coefficient of the corresponding term in the expansion in the Grassmann
variables. The spinorial superfields $W_\alpha{}^a$ and $\overline
W_\alphadot{}^a$ in Eq.~\eqref{eq:LVec}, the supersymmetric equivalents to the
field strength tensor, are related to the quantities $W_\alpha
= W_\alpha{}^a\, T_a$ and $\overline W_\alphadot = \overline W_\alphadot{}^a\,
T_a$, the latter being given by
\be
  W_\alpha= -\frac14 \bar D \cdot \bar D \ e^{2gV} D_\alpha e^{-2gV} 
  \quad  \text{and}\quad 
  \overline W_\alphadot = -\frac14  D \cdot  D \ e^{-2gV}\bar D_\alphadot e^{2gV} \ , 
\label{eq:sca2spin1}\ee
where $D$ and $\bar D$ denote the superderivatives of Eq.\
\eqref{eq:superch-and-der} and $g$ the coupling constant associated to the
gauge group. After expanding $W_\alpha$ into component fields, one obtains the
expression of $W_\alpha{}^a$,
\be 
  W_\alpha{}^a = -2g \bigg[
    -i \lambda_\alpha^a + 
    - \frac{i}{2} (\sigma^\mu \bar \sigma^\nu \theta)_\alpha V_{\mu \nu}^a 
    + \theta_\alpha D^a
    - \theta \cdot \theta (\sigma^\mu D_\mu \bar \lambda^a)_\alpha \bigg] \ ,
\label{eq:Walpha}\ee
a similar expression existing for $\overline W_\alphadot{}^a$. We recognize in
Eq.~\eqref{eq:Walpha} the expressions for the field strength tensors and the
covariant derivative in the adjoint representation
\be\bsp
  D_\mu \bar \lambda^a =&\ \del_\mu \bar \lambda^a  + g f_{bc}{}^a V_\mu^b \bar
    \lambda^c \ ,\\
  V_{\mu\nu}^a =&\ \del_\mu V_\nu^a - \del_\nu V_\mu^a + g f_{bc}{}^a V_\mu^b
    V^c_\nu \ ,
\esp \ee
where $f_{bc}{}^a$ are the structure constants of the gauge group.
In the abelian case, the expressions for the spinorial superfields can be drastically simplified, 
\be
  W_\alpha = -\frac14 \bar D \cdot \bar D D_\alpha V 
  \quad\text{and}\quad
  \overline W_{\dot \alpha}=\frac14  D \cdot  D \bar D_{\dot \alpha} V\ ,
\label{eq:sca2spin2}\ee
and the Lagrangian reduces to
\be
\lag_{U(1), V} = 
   \frac14 \big[W^\alpha W_\alpha\big]_{\theta\cdot\theta} +
   \frac14 \big[\overline W_\alphadot \overline
     W^\alphadot\big]_{\bar\theta\cdot\bar\theta} \ .
\ee

Finally, the superpotential Lagrangian $\lag_{\rm superW}$ contains the interactions among the chiral
superfields derived from the superpotential $W(\Phi)$, a holomorphic function
of the chiral superfields. It is given by
\be
   \lag_{\rm superW} = \Big[ W(\Phi) \Big]_{\theta\cdot\theta } +
    \Big[ W^\star(\Phi^\dag)\Big]_{\bar\theta\cdot\bar\theta} \ ,
\label{eq:lagsuperw}\ee
where $W^\star(\Phi^\dagger)$ is the anti-holomorphic function complex conjugate
to $W(\Phi)$. It can be shown that in a renormalizable model the superpotential can at most be cubic in the fields, thus taking the form
\be
  W(\Phi) = a_i \, \Phi^i + b_{ij}\, \Phi^i\,\Phi^j + c_{ijk}\,
    \Phi^i\,\Phi^j\,\Phi^k\ ,
\ee
for some model-dependent parameters $a_i$, $b_{ij}$ and $c_{ijk}$.

As the superpotential is simply a polynomial in the chiral superfields, it can be trivially implemented 
into \feynrules, and the Lagrangian density can easily be obtained from the 
{\tt Theta2Component} and {\tt Thetabar2\-Component} functions. The non-trivial part of any 
implementation of a supersymmetric model into \feynrules\ hence consists in the implementation of 
$\lag_{{\rm chiral}}$ and $\lag_{{\rm Yang-Mills}}$. Since these two Lagrangians are however 
completely fixed by gauge symmetry and their form is independent of the actual model under 
consideration, the superspace module of \feynrules\ comes with some predefined
functions that allow to generate $\lag_{{\rm chiral}}$ and $\lag_{{\rm
Yang-Mills}}$ in an automated way. These functions will be described in the next
section and are summarized in Table \ref{tab:autolag}.

\renewcommand{\arraystretch}{1.5}
\begin{table}
\bgfbalignshort
\multicolumn{2}{c}{\textbf{Table~\ref{tab:autolag}: Predefined functions
related to SUSY Lagrangians.}}\\
\multicolumn{2}{l}{{\tt CSFKineticTerms[csf]}}\\
& Derives all the kinetic and gauge interaction terms associated
to the chiral superfield {\tt csf}. If the function is called without any
argument, it will sum over the whole chiral content of the theory.\\
\multicolumn{2}{l}{{\tt VSFKineticTerms[vsf]}}\\
& Derives all the kinetic and gauge interaction terms associated
to the vector superfield {\tt vsf}. If the function is called without any
argument, it will sum over the whole gauge content of the theory.\\
\multicolumn{2}{l}{{\tt SuperfieldStrengthL[vsf,alpha,gaugeindex]}}\\
& Calculates the left-handed superfield strength tensor associated
to the vector superfield {\tt vsf}. The symbol {\tt alpha} denotes the
free spin index, whilst the optional symbol {\tt gaugeindex} denotes the adjoint
gauge index relevant for non-abelian gauge groups. \\
\multicolumn{2}{l}{{\tt SuperfieldStrengthR[vsf,alphadot,gaugeindex]}}\\
& Calculates the right-handed superfield strength tensor associated
to the vector superfield {\tt vsf}. The symbol {\tt alphadot} denotes the
free spin index, whilst the optional symbol {\tt gaugeindex} denotes the adjoint
gauge index relevant for non-abelian gauge groups. \\
\multicolumn{2}{l}{{\tt SolveEqMotionD[lag]}}\\
& Computes and solves the equations of motion associated to the auxiliary
$D$-fields, and then inserts the solution in the Lagrangian {\tt lag}.\\
\multicolumn{2}{l}{{\tt SolveEqMotionF[lag]}}\\
& Computes and solves the equations of motion associated to the auxiliary
$F$-fields, and then inserts the solution in the Lagrangian {\tt lag}.\\
\egfbalignshort
\textcolor{white}{\caption{\label{tab:autolag}}}
\end{table}
\renewcommand{\arraystretch}{1}

\subsection{Automatic generation of supersymmetric Lagrangians}
\label{sec:autolag}
The kinetic part of the Lagrangian for a chiral superfield $\Phi$
 can be obtained automatically in \feynrules\ from the {\tt CSFKineticTerms} function. 
 As an example, for a
chiral superfield implemented as {\tt PHI}\footnote{We assume that {\tt PHI} has been correctly declared in the \feynrules\ model file.}, the Lagrangian of Eq.\
\eqref{eq:chiralSFLag} is obtained by issuing the command
\be 
  \texttt{LChiralPhi = CSFKineticTerms[ PHI ] } \ .
\nn \ee
The expression returned by {\tt CSFKineticTerms} 
 is not automatically expanded in terms of the component
fields but still expressed in terms of superfields. 
The component-field expression of the Lagrangian can be recovered
by apply the {\tt Theta2Thetabar2Component} function to the result,
\be
  \texttt{Theta2Thetabar2Component[ LChiralPhi ] } \ .
\nn \ee

The full Lagrangian $\lag_{\rm chiral}$
is obtained by summing over all chiral superfields of the
theory. In addition to the function described above which returns the kinetic term 
for a single chiral superfield, \feynrules\ allows the user to obtain directly the complete
chiral Lagrangian expressed in terms of superfields via the command {\tt
CSFKineticTerms}.
The expression returned by this command is equivalent to a sum of terms consisting 
each in an application of the {\tt CSFKineticTerms} function to a single superfield. The extraction of the Lagrangian density can again be achieved via the {\tt Theta2Thetabar2Component} function. Hence, the full Lagrangian $\lag_{{\rm chiral}}$ can be obtained by simply issuing 
\be
  \texttt{Lchiral = Theta2Thetabar2Component[ CSFKineticTerms[ ] ]} \ .
\nn\ee

The supersymmetric equivalents of the field strength tensors can be obtained
automatically in a similar way. The left-handed superfield strength tensors
$W_\alpha$ and $W_\alpha{}^a$ associated to a vector superfield {\tt V} can be
called in the superspace module of \feynrules\ via the commands\footnote{We
assume that the vector superfield {\tt V} has been declared associated to some
gauge group in the \feynrules\ model file.}
\be\bsp
  &\ \texttt{SuperfieldStrengthL [ V, sp ] }  \ ,\\
  &\ \texttt{SuperfieldStrengthL [ V, sp, ga] } \ ,
\esp \nn \ee
in the abelian and non-abelian cases. The symbol {\tt
sp} denotes the undotted spin index attached to the spinorial
superfield while {\tt ga} is the adjoint gauge index relevant for
non-abelian gauge groups. Similarly, the abelian and non-abelian right-handed
superfield strength tensors $\overline W_\alphadot$ and $\overline
W_\alphadot{}^a$ can be obtained through
\be\bsp
  &\ \texttt{SuperfieldStrengthR [ V, sp ] }  \ ,\\
  &\ \texttt{SuperfieldStrengthR [ V, sp, ga ] } \ ,
\esp \nn \ee
respectively, the only difference with the left-handed case being the variable
{\tt sp} which stands this time for a dotted spin index. Note that
the spinorial superfields defined by Eq.\
\eqref{eq:sca2spin1} and Eq.\ \eqref{eq:sca2spin2} are not hardcoded in the
superspace module of \feynrules, and will be recalculated each time. However,
the {\tt SuperfieldStrengthL} and {\tt SuperfieldStrengthR}
functions will be evaluated by \feynrules\ only at the time of the expansion in
terms of the component fields.

From the superfield strength tensors we can easily built the kinetic terms for vector superfields in an automated way. This is achieved in \feynrules\ by issuing the command
\be
 \texttt{LV = VSFKineticTerms [ V ] } \ ,
\nn \ee
The Super-Yang-Mills Lagrangian of Eq.\ \eqref{eq:LVec} can then easily be obtained by extracting the $\theta\cdot\theta$ and $\bar\theta\cdot\bar\theta$ components,
\be
\texttt{LSYM = Theta2Component[ LV ] + Thetabar2Component[ LV ]} \ .
 \nn \ee
If a model contains several gauge groups, we have to sum over the corresponding
kinetic terms. Similarly to the automatic generation of the kinetic terms of
chiral superfields, issuing {\tt VSFKineticTerms} without any argument is
equivalent to a sum over all possible vector superfields defined in the model.

At this stage, generating a Lagrangian density for any supersymmetric model reduces to
an almost trivial task with the help of the functions that we have just
described. Assuming that a superpotential {\tt SP} has been defined, the full
Lagrangian density can be easily implemented into \feynrules\ as
\be \bsp
  \texttt{Lag =}&\ \texttt{Lchiral + LSYM + LW}  \ ,
\esp \nn\ee
where the terms in the sum in the right-hand side are given by
\be \bsp
  \texttt{LC =}&\ \texttt{CSFKineticTerms[ ] } \ ,\\
  \texttt{Lchiral =}&\ \texttt{Theta2Thetabar2Component[ LC ] } \ ,\nn\\
  \texttt{LV =}&\ \texttt{VSFKineticTerms[ ]} \ , \\ 
  \texttt{LSYM =}&\ \texttt{Theta2Component[LV] + Thetabar2Component[LV]} \ , \\
  \texttt{LW =}&\ \texttt{Theta2Component[SP]+Thetabar2Component[HC[SP]]} \ .
\esp\ee
The Lagrangian density obtained in this way however still depends on the
auxiliary $F$ and $D$ fields, which can be eliminated by their equations of motion. 
This can be performed automatically with the help of
two functions {\tt SolveEqMotionD} and {\tt
SolveEqMotionF}. Each of them computes, for the $D$-fields and $F$-fields respectively,
the equations of motion directly from the Lagrangian, solves them analytically
and subsequently inserts the solution into the Lagrangian in order to eliminate the
auxiliary fields. 
Using the Lagrangian defined previously, the auxiliary fields are
eliminated via the \mathematica\ commands
\begin{verbatim}
    Lag = SolveEqMotionD[Lag] ,
    Lag = SolveEqMotionF[Lag] .
\end{verbatim}

\section{Implementation of the Minimal Supersymmetric Standard Model}
\label{sec:mssm}
The Minimal Supersymmetric Standard Model (MSSM) is the simplest supersymmetric
model, resulting from a direct supersymmetrization of the Standard Model (SM)
\cite{Nilles:1983ge, Haber:1984rc}. In this section, we describe its 
implementation in \feynrules\ in terms of superfields. The corresponding model file
can be downloaded from the \feynrules\ website:\\
$~~~~$ {\tt http://feynrules.phys.ucl.ac.be/wiki/MSSM}.

\subsection{Gauge groups and representations} \label{sec:mssmgauge}
The MSSM is based on the same gauge group as the SM,
$SU(3)_c \times SU(2)_L \times U(1)_Y$. The implementation of these
three gauge group classes slightly differs from the one included into the
previous MSSM implementation, expressed entirely in terms of the
component fields \cite{Christensen:2009jx}. First, we associate to each gauge
group one vector superfield instead of one gauge boson. The abelian factor
$U(1)_Y$ is hence implemented as
\begin{verbatim}
  U1Y  == { 
    Abelian          -> True,
    CouplingConstant -> gp,
    Superfield       -> BSF, 
    Charge           -> Y
  },
\end{verbatim}
where the vector superfield {\tt BSF} will be specified below.
Secondly, the implementation of the non-abelian direct factors of the gauge
group, $SU(2)_L$ and $SU(3)_c$, includes a consistent definition of the
representation matrices, together with the associated
index type, related to the representations in which
one or several chiral superfields of the model lie\footnote{We let \feynrules\
handle automatically the adjoint representations necessary for the vector
superfields of the model. We refer to the \feynrules\ manual for more
information.}. This allows to
extract the gauge interaction and kinetic terms of the Lagrangian automatically
with the help of the two functions {\tt CSFKinetic\-Terms} and {\tt
VSFKineticTerms}. 
For $SU(2)_L$, we only need the fundamental representation, labelled by the
quantity {\tt Ta}, and defined together with the associated gauge index {\tt
SU2D}. This is implemented as,
\begin{verbatim}
  SU2L == { 
    Abelian           -> False, 
    CouplingConstant  -> gw, 
    Superfield        -> WSF, 
    StructureConstant -> ep, 
    Representations   -> {Ta,SU2D}, 
    Definitions       -> {
         Ta[a__]        -> PauliSigma[a]/2, 
         ep             -> Eps}
  }  .
\end{verbatim}
For the $SU(3)_c$ gauge group, both the fundamental and its complex conjugate 
representation are needed, the gauge group being hence defined as,
\begin{verbatim}
  SU3C == 
  { 
    Abelian           -> False,
    CouplingConstant  -> gs, 
    DTerm             -> dSUN,
    Superfield        -> GSF,
    StructureConstant -> f,
    Representations   -> { {T,Colour}, {Tb,Colourb} }
  } ,
\end{verbatim}
where {\tt Colour} and {\tt Colourb} are the fundamental and
antifundamental representation indices and {\tt T} and {\tt Tb} 
the corresponding representation matrices. It is important to note
that all
antifundamental indices in colour space must be replaced by fundamental
ones before exporting the Feynman rules to the interfaces to Monte Carlo
generators, following the conventions of Ref.~\cite{Christensen:2008py}.

\subsection{Field content}
The Standard Model quarks and leptons are embedded into chiral supermultiplets,
together with their squark and slepton partners, which are
grouped into three generations of six chiral superfields,
\be\bsp
 &\ Q_L^i = ({\utilde {\bf 3}}, {\utilde{\bf 2}}, \frac16) \quad  ,\quad
    U_R^i = ({\utilde {\bf \bar 3}}, {\utilde{\bf 1}},-\frac23) \quad ,\quad
    D_R^i = ({\utilde {\bf \bar 3}}, {\utilde{\bf 1}}, \frac13) \\
 &\ L_L^i = ({\utilde {\bf 1}}, {\utilde{\bf 2}},-\frac12) \quad , \quad
    E_R^i = ({\utilde {\bf 1}}, {\utilde{\bf 1}}, 1) \quad , \quad 
    V_R^i = ({\utilde {\bf 1}}, {\utilde{\bf 1}}, 0) \ ,
\esp\label{eq:SFnames}\ee
where $i$ stands for a generation index and where we have indicated the
representations of the different superfields under the MSSM gauge
group\footnote{It can then be seen that the representation matrices defined in
Section \ref{sec:mssmgauge} are enough to describe the entire superfield content
given in Eq.\ \eqref{eq:SFnames}.}. The component fields included in each
superfield can be found in Table \ref{tab:chiral1}. For completeness, 
the right-handed neutrino superfield
has been introduced, but it will be kept sterile, \ie, non-interacting with any
other superfield.
\renewcommand{\arraystretch}{1.1}
\begin{table}[!t]
  \begin{center}
  \begin{tabular}{|c||c|c|c|c|}
    \hline
    Superfield & \begin{tabular}{c}Standard Model\\ fermion\end{tabular} & 
     Superpartner & Representation \\ \hline 
    \multirow{4}{*}{$Q_L^i$}&&&\\
       &$q_L^i = \bpm u_L^i \\ d_L^i \epm$ & 
       $\tilde q^i_L = \bpm \tilde u^i_L\\  \tilde d^i_L\epm$ &
       $({\utilde {\bf 3}}, {\utilde{\bf 2}}, \frac16)$ \\&&&\\
    $U^i_R$ & 
       $u_R^{ic}$ & 
       $\tilde u_R^{i\dag}$ & 
       $({\utilde{\bf \bar 3}},{\utilde{\bf 1}}, -\frac23)$\\
    $D_R^i$ &
       $d_R^{ic}$ & 
       $\tilde d_R^{i\dag}$ & 
       $({\utilde{\bf \bar 3}},{\utilde{\bf 1}}, \frac13)$\\&&&\\
    \hline
    \multirow{4}{*}{$L_L^i$}&&&\\
       &$\ell_L^i = \bpm \nu_L^i \\ e_L^i\epm$ & 
       $\tilde \ell_L^i = \bpm\tilde \nu^i_L\\  \tilde e^i_L\epm$ &
       $({\utilde{\bf 1}}, {\utilde{\bf 2}}, -\frac12)$\\&&&\\
    $E_R^i$&
       $e_R^{ic}$ & 
       $\tilde e_R^{i\dag}$ & 
       $({\utilde{\bf 1}},{\utilde{\bf 1}}, 1)$\\
    $V_R^i$&
       $\nu_R^{ic}$ & 
       $\tilde \nu_R^{i\dag}$ & 
       $({\utilde{ \bf 1}},{\utilde{\bf 1}}, 0)$\\&&&\\
    \hline 
  \end{tabular}
  \end{center} 
  \caption{\label{tab:chiral1}The MSSM (s)fermion sector, with the
    representations under the gauge group $SU(3)_c \times SU(2)_L \times
    U(1)_Y$. The superscript $c$ denotes charge conjugation. }
\end{table}
\renewcommand{\arraystretch}{1}

The six chiral superfields of Eq.\ \eqref{eq:SFnames} are implemented
following the instructions given in Section \ref{sec:Csuperfields}, while the
component fields are implemented following the synthax presented in Refs.\
\cite{Christensen:2008py, Butterworth:2010ym} and the constraints introduced in
Section \ref{sec:Csuperfields}. Hence, for each chiral superfield and the
associated component fields, the
hypercharge quantum number and the attached gauge indices are specified. 
This allows the automatic function {\tt CSFKineticTerms} to correctly derive
the associated gauge interactions. As an example, the 
weak isospin doublet of quarks $Q_L^i$ is implemented using the
\mathematica\ instructions 
\begin{verbatim}
  CSF[1] == { 
    ClassName      -> QL, 
    Chirality      -> Left, 
    Weyl           -> QLw,  
    Scalar         -> QLs,  
    QuantumNumbers -> {Y-> 1/6}, 
    Indices        -> {Index[SU2D],Index[GEN],Index[Colour]}} .
\end{verbatim}
In our model implementation, we follow a simple naming
scheme for the component fields where the names of the Weyl fermionic and scalar
components are obtained by suffixing {\tt w} and {\tt s} to the superfield class
name, respectively.

To preserve the electroweak symmetry from chiral anomalies and in order to give
masses to both up-type and down-type fermions, the MSSM Higgs sector 
contains two chiral supermultiplets,
\be\bsp
  H_D = ({\utilde{\bf 1}}, {\utilde{\bf 2}}, -\frac12)\quad ,\quad
  H_U = ({\utilde{\bf 1}}, {\utilde{\bf 2}},  \frac12) \ ,
\esp\label{eq:SFHiggs}\ee
each of which consisting in one scalar Higgs $SU(2)_L$ doublet and its fermionic Higgsino
partner, as shown in Table \ref{tab:chiral2}. The superfield
$H_U$ couples to up-type particles whilst $H_D$ couples to down-type particles.
The two chiral superfields $H_U$ and $H_D$ are implemented in a similar fashion as presented
above.

\begin{table}[!t]
  \begin{center}
  \begin{tabular}{|c||c|c|c|c|}
    \hline
    Superfield & Higgs boson & Higgsino & Representation \\ \hline 
    \hline 
    \multirow{4}{*}{$H_D$}&&&\\
       &$H_d = \bpm H_d^0 \\ H_d^- \epm$&
        $\widetilde H_d = \bpm \widetilde H_d^0 \\ \widetilde H_d^- \epm$&
        $({\utilde{\bf 1}}, {\utilde{\bf 2}}, -\frac12)$ \\&&&\\
    \multirow{4}{*}{$H_U$}&&&\\
       &$H_u = \bpm H_u^+ \\ H_u^0 \epm$ & 
        $\widetilde H_u = \bpm \widetilde H_u^+ \\ \widetilde H_u^0\epm$&
        $({\utilde{\bf 1}}, {\utilde{\bf 2}}, \frac12)$ \\&&&\\
\hline
\end{tabular}
  \end{center} 
  \caption{\label{tab:chiral2}The MSSM Higgs(ino) sector, with the
    representations under the gauge group $SU(3)_c \times SU(2)_L \times U(1)_Y$.}
\end{table}
The gauge sector is described by three vector superfields associated each
to one specific direct factor of the gauge group. They lie in the
corresponding adjoint representation and are singlets under all the other group
factors,
\be \bsp
SU(3)_c \to &\ \Phi_G = ({\utilde{\bf 8}}, {\utilde{\bf 1}}, 0)\ ,\\
SU(2)_L \to &\ \Phi_W = ({\utilde{\bf 1}}, {\utilde{\bf 3}}, 0)\ ,\\
 U(1)_Y \to &\ \Phi_B = ({\utilde{\bf 1}}, {\utilde{\bf 1}}, 0)\ .
\esp\ee
\renewcommand{\arraystretch}{1.2}
\begin{table}[!t]
  \begin{center}
  \begin{tabular}{|c||c|c|c|c|}
    \hline
    Superfield & Gauge boson & Gaugino & Representation \\ \hline 
    \hline 
    $\Phi_B$ & $B_\mu$ & $\widetilde B$ & $({\utilde{\bf 1}}, {\utilde{\bf 1}}, 0)$ \\
    $\Phi_W$ & $W_\mu$ & $\widetilde W$ & $({\utilde{\bf 1}}, {\utilde{\bf 3}}, 0)$ \\
    $\Phi_G$ & $g_\mu$ & $\widetilde g$ & $({\utilde{\bf 8}}, {\utilde{\bf 1}}, 0)$ \\
    \hline
  \end{tabular}
  \end{center} 
  \caption{\label{tab:gauge}The MSSM gauge sector, with the
    representations under the gauge group $SU(3)_c \times SU(2)_L \times U(1)_Y$.}
\end{table}
\renewcommand{\arraystretch}{1}
These superfields include, in addition to the Standard Model gauge bosons,
their fermionic partners, the gauginos, as shown in Table \ref{tab:gauge}.
Each vector superfield is implemented following the same pattern, with
the {\tt Indices} option set to the adjoint index of the relevant gauge group,
labelled by {\tt SU2W} and {\tt Gluon} for $SU(2)_L$ and $SU(3)_c$,
respectively. As an example, the gluon superfield implementation is given by
\begin{verbatim}
  VSF[3] == { 
    ClassName  -> GSF, 
    GaugeBoson -> G,
    Gaugino    -> gow,
    Indices    -> {Index[Gluon] }
  }  .
\end{verbatim}

\subsection{Lagrangian} \label{sec:mssmlag}
As stated in Section \ref{sec:lagrangians}, the kinetic and gauge interaction
terms of the chiral and vector superfields are entirely fixed by gauge
invariance. For the MSSM, these terms read, 
\be\bsp
  \lag_{\rm SYM} = &\ \bigg[
    \frac{W_B^\alpha W_{B\alpha}}{4} + 
    \frac{W_{Wk}^\alpha W^k_{W\alpha}}{16 g_w^2}  + 
    \frac{W_{Ga}^\alpha W^a_{G\alpha}}{16 g_s^2} 
   \bigg]_{\theta\cdot\theta} + \hc \ ,\\
 \lag_{\rm chiral} = &\ \bigg[
    Q_L^\dag \Big(e^{-\frac13 g^\prime \Phi_B} e^{-2 g_w V_W} e^{-2 g_s V_G}\Big)
      Q_L \ + \\
  &\quad
    U_R^\dag \Big(e^{ \frac43 g^\prime \Phi_B} e^{-2 g_s V^\prime_G} \Big) U_R \ + \ 
    D_R^\dag \Big(e^{-\frac23 g^\prime \Phi_B} e^{-2 g_s V^\prime_G} \Big) D_R \ + \\
  &\quad
    L_L^\dag \Big(e^{ g^\prime \Phi_B} e^{-2 g_w V_W} \Big) L_L  \ + \ 
    E_R^\dag \Big(e^{-2 g^\prime \Phi_B} \Big) E_R  \ + \
    V_R^\dag V_R  \ + \\
  &\quad
    H_D^\dag \Big(e^{ g^\prime \Phi_B} e^{-2 g_w V_W} \Big) H_D  \ + \
    H_U^\dag \Big(e^{-g^\prime \Phi_B} e^{-2 g_w V_W} \Big) H_U 
  \bigg]_{\theta\cdot\theta \bar\theta\cdot\bar\theta}  \ ,
\esp\label{eq:MSSMCV}\ee
where
\be\bsp
  W_{B\alpha} =&\ -\frac14 \bar D \cdot \bar D D_\alpha \Phi_B \ , \\
  W_{W\alpha} =&\ -\frac14 \bar D \cdot \bar D e^{2 g_w V_W} D_\alpha e^{-2 g_w
    V_W} \quad \text{and}\quad  W_{W\alpha} = W_{W\alpha}^k \frac12\sigma_k \ , \\
  W_{G\alpha} =&\ -\frac14 \bar D \cdot \bar D e^{2 g_s V_G} D_\alpha e^{-2 g_s
    V_G} \quad \text{and}\quad  W_{G\alpha} = W_{G\alpha}^a T_a \ .
\esp\ee
In the expressions above, we have introduced the non-abelian vector superfields
$V_W = \Phi_{W^k} \frac12 \sigma_k$, $V_G = \Phi_{G^a} T_a$ and $V_G^\prime =
\Phi_{G^a} \bar T_a$ where $\sigma^k/2$ denote the generators of the fundamental
representation of $SU(2)_L$, $\sigma_k$ being the Pauli matrices, and where
$T_a$ and $\bar T_a$ are the generators of the fundamental and antifundamental
representation of $SU(3)_c$. The gauge coupling constants are defined
as $g^\prime$, $g_w$ and $g_s$ and all generation indices are understood.
The two Lagrangians appearing in Eq.\ \eqref{eq:MSSMCV} are implemented in the
model file as described in Section \ref{sec:autolag}, using the automatized
functions {\tt CSFKineticTerms} and {\tt VSFKineticTerms}.

The interactions among the chiral superfields introduced in Eq.\
\eqref{eq:SFnames} and Eq.\ \eqref{eq:SFHiggs} are included in the
superpotential
\be \bsp
  W_{\rm MSSM} = &\ 
    ({\bf y^u})_{ij} U_R^i Q_L^j \!\cdot\! H_U - 
    ({\bf y^d})_{ij} D_R^i Q_L^j \!\cdot\! H_D - 
    ({\bf y^e})_{ij} E_R^i L_L^j \!\cdot\! H_D \\
  &\ + \mu H_U \!\cdot\! H_D \ , 
\esp \label{eq:wmssmold}\ee
where ${\bf y^u}$, ${\bf y^d}$ and ${\bf y^l}$ denote the $3\times3$ Yukawa
matrices in flavor space, $\mu$ the Higgs off-diagonal mass-mixing and the dot
products stand for $SU(2)$ invariant products. This superpotential is the
most general function satisfying renormalizability, the gauge symmetries of the model
and $R$-parity conservation. Our conventions regarding the parameters of the
model follow the SUSY Les Houches Accord (SLHA) \cite{Skands:2003cj,
Allanach:2008qq} and the Yukawa matrices must then be given flavor-diagonal. However,
the fermionic components of the superfields given in Eq.\ \eqref{eq:SFnames} are
gauge-eigenstates and not mass-eigenstates. In the model file, we address this
issue by implementing the modified superpotential
\be \bsp
  W_{\rm MSSM} = &\ 
    ({\bf \hat y^u})_{ij} U_R^i Q_L^j \!\cdot\! H_U - 
    ({\bf \hat y^d} V_{\rm CKM}^\dag)_{ij} D_R^i Q_L^j \!\cdot\! H_D - 
    ({\bf \hat y^e})_{ij} E_R^i L_L^j \!\cdot\! H_D \\
  &\ + \mu H_U \!\cdot\! H_D \ , 
\esp \label{eq:wmssm} \ee
where the hatted Yukawa matrices are flavor-diagonal, following thus the SLHA
conventions, and where the superfields are gauge-eigenstates. All the rotations
diagonalizing the gauge-eigenstate basis will be absorbed at the
component field level rather than at the superfield level, as described in
Section \ref{sec:MSSMewsb}, and the role of the CKM matrix
appearing in the second term of the superpotential is to compensate the only
remaining misalignment between mass and gauge-eigenstates after
these rotations. 
The Lagrangian associated to this superpotential is given by Eq.\ \eqref{eq:lagsuperw}. 

Finally, the supersymmetry-breaking Lagrangian is obtained by adding
explicitly all soft supersymmetry-breaking terms at low-energy, 
\be \bsp 
   \lag_{\rm Soft} =
   &\ - {1 \over 2} \Big[ 
     M_1\ \widetilde B \!\cdot\! \widetilde B + 
     M_2\ \widetilde W \!\cdot\! \widetilde W + 
     M_3\ \widetilde g \!\cdot\! \widetilde g + 
     \hc \Big]\\
   &\ - ({\bf m^2_{\tilde Q}})^i{}_j \tilde q_{Li}^\dag\ \tilde q_L^j 
      - ({\bf m^2_{\tilde U}})^i{}_j \tilde u_{Ri}\ \tilde u_R^{j\dag} 
      - ({\bf m^2_{\tilde D}})^i{}_j \tilde d_{Ri}\ \tilde d_R^{j\dag} \\
   &\ - ({\bf m^2_{\tilde L}})^i{}_j \tilde \ell_{Li}^\dag\ \tilde \ell_L^j 
      - ({\bf m^2_{\tilde E}})^i{}_j \tilde e_{Ri}\ \tilde e_R^{j\dag} 
      - m_{H_u}^2 H_u^\dag\ H_u  
      - m_{H_d}^2 H_d^\dag\ H_d\\
   &\ + \Big[ - ({\bf T^u})_{ij} \tilde u_R^{i\dag} \tilde q_L^j \!\cdot\! H_u 
              + ({\bf T^d})_{ij} \tilde d_R^{i\dag} \tilde q_L^j \!\cdot\! H_d 
              + ({\bf T^e})_{ij} \tilde e_R^{i\dag} \tilde \ell_L^j \!\cdot\!
                H_d\\
   & \qquad - b H_u \!\cdot\! H_d  + \hc \Big]\ .
\esp \ee
The first line of this equation contains the gaugino mass terms, the second and
third lines the scalar mass terms, ${\bf m^2_{\tilde Q}}$, ${\bf
m^2_{\tilde L}}$, ${\bf m^2_{\tilde u}}$, ${\bf m^2_{\tilde d}}$, ${\bf
m^2_{\tilde e}}$ being $3\times3$ hermitian matrices in generation space and the
fourth and fifth line the bilinear and trilinear scalar interactions derived
from the superpotential. ${\bf T_u}$, ${\bf T_d}$, and ${\bf T_e}$ are
$3\times 3$ matrices in generation space. Similarly to the superpotential of Eq.\
\eqref{eq:wmssm}, we write the scalar trilinear interactions as,
\be
  \Big[ 
  - ({\bf \hat T^u})_{ij} \tilde u_R^{i\dag} \tilde q_L^j \!\cdot\! H_u 
  + ({\bf \hat T^d} V_{\rm CKM}^\dag)_{ij} 
    \tilde d_R^{i\dag} \tilde q_L^j \!\cdot\! H_d 
  + ({\bf \hat T^e})_{ij} \tilde e_R^{i\dag} \tilde \ell_L^j \!\cdot\! H_d + \hc \Big]
\ee
rather than in their original form, following thus the SLHA conventions. 
Finally, the equation of motions for the auxiliary fields are solved so that they
are eliminated from the Lagrangian as described in Section \ref{sec:autolag}.

\subsection{Electroweak symmetry breaking, particle mixings and Dirac fermions}
\label{sec:MSSMewsb}
Due to the strong Yukawa coupling between the superfields $H_U$, $Q_L^3$ and
$U_R^3$ in the superpotential, the electroweak symmetry is radiatively broken to
electromagnetism and the classical Higgs potential has a non-trivial minimum.
Shifting the neutral scalar Higgs bosons by their vacuum expectation values (vevs), 
\be
  H_u^0 \to  {v_u + h_u^0 \over \sqrt{2}} \quad \text{and} \quad  
  H_d^0 \to  {v_d + h_d^0 \over \sqrt{2}} \ ,
\label{eq:vev}\ee 
where $v_u$ and $v_d$ denote the two vevs of the neutral 
Higgs bosons and $h_u^0$ and $h_d^0$ complex scalar fields, 
we can extract the mass matrices of the electroweak gauge bosons $B_\mu$ and
$W_\mu^k$, diagonalize them, and derive the physical mass-eigenstates, the
photon $A_\mu$ and the weak bosons $W_\mu^\pm$ and $Z_\mu$. As in the Standard
Model, the transformation rules relating the mass and interaction bases are
\be
  W_\mu^\pm = {1\over \sqrt{2}} (W_\mu^1 \mp i W^2_\mu),\quad \text{and} \quad
  \bpm  Z_\mu \\ A_\mu \epm = \bpm \cos\theta_w & -\sin\theta_w\\ \sin\theta_w
    & \cos\theta_w \epm \bpm W_\mu^3 \\ B_\mu \epm,
\ee
where the weak mixing angle $\theta_w$ and the physical masses $M_Z$ and $M_W$
are defined by
\be
   \cos^2\theta_w = {g_w^2 \over g_w^2+g^{\prime 2}} \ , \quad
   M_Z = {g_w v \over 2 \cos\theta_w} \quad \text{and} \quad
   M_W = {g_w v \over 2} \ ,
\ee
with $v^2 = v_u^2+v_d^2$.
We define consistently all the mixing parameters in the model file, taking the
masses as input parameters. The mixing angles and the vevs are then dependent on
the latter.
The rotations are implemented using the {\tt Definitions} option of the
particle class. As an example, the $SU(2)_L$ boson redefinitions are
implemented as 
\begin{verbatim}
  Definitions -> {
    Wi[mu_,1] -> (Wbar[mu]+W[mu])/Sqrt[2], 
    Wi[mu_,2] -> (Wbar[mu]-W[mu])/(I*Sqrt[2]), 
    Wi[mu_,3] -> cw Z[mu] + sw A[mu]
  } ,
\end{verbatim}
where {\tt A}, {\tt Z} and {\tt W} correspond to the model file definitions of
the physical gauge bosons, containing all the options required in
order to have the interfaces to the Monte Carlo codes and to \feynarts\ working
properly ({\tt PDG}, {\tt PropagatorType}, ...), as presented in Ref.\
\cite{Christensen:2008py}.

In the Higgs sector, three out of the eight real degrees of freedom of the two
doublets are the pseudo-Goldstone bosons $G^\pm$ and $G^0$ becoming the
longitudinal modes of the weak bosons, while the five others mix to the physical
Higgses, $h^0$, $H^0$, $A^0$ and $H^\pm$. The diagonalization of the scalar,
pseudoscalar and charged Higgs mass matrices leads to the transformation rules
\be\bsp
  h_u^0 =&\  \cos\alpha\ h^0 + \sin\alpha\ H^0 + 
    i \cos\beta\ A^0 + i \sin\beta\ G^0\ , \\ 
  h_d^0 =&\ -\sin\alpha\ h^0 + \cos\alpha\ H^0 + 
    i \sin\beta\ A^0 - i \cos\beta\ G^0\ , \\
  H_u^+ =&\  \cos\beta\ H^+ +\sin\beta\ G^+\ , \\  
  H_d^- =&\  \sin\beta\ H^- - \cos\beta\ G^-\ , 
\esp\label{eq:HiggsMixing}\ee 
where $\alpha$ is the neutral Higgs mixing angle and the $\beta$ angle
is defined by $\tan\beta=v_u/v_d$. Following the SLHA
conventions \cite{Skands:2003cj, Allanach:2008qq}, all the mixing angles are
external parameters included in the SLHA block {\tt HMIX}, and the rotations are
implemented in the scalar Higgs field class definition, \eg, as 
\begin{verbatim}
  Definitions -> { 
    hus[1] -> Cos[beta]*H + Sin[beta]*GP, 
    hus[2] -> (vu + Cos[alp]*h0 + Sin[alp]*H0 + 
       I*Cos[beta]*A0 + I*Sin[beta]*G0)/Sqrt[2]  
  } 
\end{verbatim}
for the $H_u$ doublet labelled by {\tt hus}, 
where {\tt H}, {\tt A0}, {\tt h0}, {\tt H0}, {\tt GP} and {\tt G0} are the labels
of the physical Higgs and Goldstone bosons. The latter are implemented following the
synthax presented in Ref.\ \cite{Christensen:2008py} and contain again all
the options required by the interfaces to work properly.

In the fermionic sector, the mass matrix of the neutral partners of the gauge
and Higgs bosons is diagonalized through a unitary matrix $N$ which relates the
four physical (two-component) neutralinos $\chi^0_i$ to the
interaction-eigenstates,
\be 
  \bpm \chi^0_1 \\ \chi^0_2 \\ \chi^0_3 \\ \chi^0_4 \epm = N 
  \bpm i \widetilde B \\ i \widetilde W^3 \\ \widetilde H_d^0 \\ 
     \widetilde H_u^0 \epm \ .
\label{eq:neutralinos}\ee 
Similarly, the mass matrix of the charged partners is diagonalized through two
unitary matrices $U$ and $V$ relating the interaction-eigenstates to the
physical (two-component) charginos eigenstates $\chi^\pm_i$ according to
\be
  \bpm \chi^+_1 \\ \chi^+_2 \epm = V \bpm i \widetilde W^+ \\ \widetilde H^+_u \epm 
  \quad \text{and}\quad  
  \bpm \chi^-_1 \\ \chi^-_2 \epm = U \bpm i \widetilde W^- \\ \widetilde H^-_d
    \epm \ ,
\label{eq:charginos}\ee
where the charged winos have undergone the same rotations as the charged gauge
bosons,
\be
  \widetilde W_\mu^\pm = {1\over \sqrt{2}} (\widetilde W_\mu^1 \mp i \widetilde
    W^2_\mu) \ . 
\ee
The factors of $i$ absorbed in the gaugino definition are required in order to
obtain real mass matrices. These conventions may seem a bit
different from the ones specified by the SUSY Les Houches Accord, accounting for
factors of $-i$. However, this sign is purely conventional and is related to the
sign of the exponentials in the Lagrangian of Eq.\ \eqref{eq:MSSMCV}. The
choices of Eq.\ \eqref{eq:neutralinos} and Eq.\ \eqref{eq:charginos} ensures
that the mixing matrices $N$, $U$ and $V$ are the same as those specified in the
accord.
In the MSSM model file, the real and imaginary parts of these mixing
matrices are considered as input parameters, following the SLHA, while
the full matrices themselves are dependent parameters. The physical two-component
neutralino fields are implemented following the synthax of Ref.\
\cite{Butterworth:2010ym} and the rotations are included in the {\tt
Definitions} option of the gauge-eigenstate particle classes. As an example, the
wino rotations are given by
\begin{verbatim}
  Definitions -> {
    wow[s_,1] :> Module[{i}, (Conjugate[UU[i,1]]*chmw[s,i] + 
        Conjugate[VV[i,1]]*chpw[s,i])/(I*Sqrt[2]) ], 
    wow[s_,2] :> Module[{i}, (Conjugate[UU[i,1]]*chmw[s,i] - 
        Conjugate[VV[i,1]]*chpw[s,i])/(-Sqrt[2]) ], 
    wow[s_,3] :> Module[{i}, -I*Conjugate[NN[i,2]]*neuw[s,i] ] 
  } ,
\end{verbatim}
where {\tt chmw}, {\tt chpw} and {\tt neuw} denote the labels of the physical
$\chi^-$, $\chi^+$ and $\chi^0$ fields, respectively, and {\tt NN}, {\tt UU} and
{\tt VV} the mixing matrices.

The diagonalization of the quark sector requires four unitary matrices,
\be
  d_L^i \to V_d d_L^i \ , \quad 
  d_R^{ic} \to U_d d_R^{ic}\ , \quad 
  u_L^i \to V_u u_L^i\ , \quad 
  u_R^{ic} \to U_u u_R^{ic} \ , 
\ee
so that the superpotential of Eq.\ \eqref{eq:wmssmold} is rotated to a form
where the Yukawa matrices are diagonal. As a consequence, the charged current
weak interactions become proportional to the CKM matrix
\be
  V_{\rm CKM} = V_u^\dag V_d \ .
\ee 
We adopt the standard choice of absorbing these rotations in a redefinition of the
down-type quark fields alone,
\be
  d_L^i \to V_{\rm CKM} d_L^i \ ,
\ee
keeping the up-type quark fields unchanged. This field redefinition is implemented in
\feynrules\ through the {\tt Definitions} option of the {\tt QLw} class,
\begin{verbatim}
  Definitions -> {
    QLw[s_, 1, ff_, cc_] -> uLw[s,ff,cc], 
    QLw[s_, 2, ff_, cc_] :> Module[{ff2}, 
       CKM[ff,ff2] dLw[s,ff2,cc] ]
  }  ,
\end{verbatim}
where {\tt uLw} and {\tt dLw} denote the (two-component) quark mass eigenstates
and the CKM matrix is included in the model file through the SLHA blocks {\tt
VCKM} and {\tt IMVCKM} for its real and imaginary parts, respectively, the
complete matrix being thus an internal parameter.
Similarly, the lepton sector is diagonalized through the rotations,
\be
  e_L^i \to V_e e_L^i \ , \quad 
  e_R^{ic} \to U_e e_R^{ic}\ , \quad 
  \nu_L^i \to V_\nu \nu_L^i\ , 
\ee
rendering the charged current interactions proportional to the PMNS
matrix
\be
  V_{\rm PMNS} = V_e^\dag V_\nu \ .
\ee
We adopt here the choice of absorbing the rotations in a redefinition of the
neutrino fields alone,
\be
  \nu_L^i \to V_{\rm PMNS} \nu_L^i \ ,
\ee
leaving the charged lepton fields unchanged\footnote{There are many different
models accounting for neutrino masses. Since there is thus not any unique way to
define neutrino mass terms, we are not including them in the implemented
Lagrangian and we only account for neutrino mixings through field
redefinitions.}.
The fermion rotations presented above legitimate the implementation of the
superpotential under the form of Eq.\ \eqref{eq:wmssm}. It can be checked that
the extracted quark and lepton mass matrices are indeed diagonal and no further
rotation is necessary.
The PMNS matrix is implemented in the model
file within the SLHA blocks {\tt UPMNS} and {\tt IMUPMNS} for its real and
imaginary parts, respectively, while again, the entire matrix is considered as
an internal parameter. On a similar fashion as for the quark sector, the
implementation of the field redefinitions is given by
\begin{verbatim}
  Definitions -> {
    LLw[s_, 1, ff_] :> Module[{ff2}, 
      PMNS[ff,ff2]*vLw[s,ff2] ], 
    LLw[s_,2,ff_]->eLw[s,ff]
  }  ,
\end{verbatim}
where {\tt LLw} is the gauge-eigenstate field and {\tt vLw} and {\tt eLw} the
mass-eigenstate ones, the right-handed components being unchanged.

In the scalar sector, we define the super-CKM and super-PMNS bases as the bases in
which the scalar fields undergo the same rotations as their fermionic
counterparts. However, the fermion and sfermion fields can be misaligned due to
possible off-diagonal mass terms in the supersymmetry-breaking Lagrangian
$\lag_{\rm Soft}$, and four additional rotations $R^u$, $R^d$, $R^e$ and $R^\nu$
are (in general) required,
\be\bsp 
  \bpm \tilde u_1\\ \tilde u_2\\ \tilde u_3\\ \tilde u_4\\ \tilde u_5 \\ \tilde
    u_6 \epm = R^u \bpm \tilde u_L\\ \tilde c_L\\ \tilde t_L\\ \tilde u_R \\
    \tilde c_R\\ \tilde t_R \epm \ ,\quad
  \bpm \tilde d_1\\ \tilde d_2\\ \tilde d_3\\ \tilde d_4\\ \tilde d_5 \\ \tilde
    d_6 \epm =&\ R^d \bpm \tilde d_L\\ \tilde s_L\\ \tilde b_L\\ \tilde d_R \\
    \tilde s_R\\ \tilde b_R \epm \ ,\quad
  \bpm \tilde e_1\\ \tilde e_2\\ \tilde e_3\\ \tilde e_4\\ \tilde e_5 \\ \tilde
    e_6 \epm = R^e \bpm \tilde e_L\\ \tilde \mu_L\\ \tilde \tau_L\\ \tilde e_R \\
    \tilde \mu_R\\ \tilde \tau_R \epm \ , \\
 \bpm \tilde \nu_1\\ \tilde \nu_2\\ \tilde \nu_3 \epm = &\ R^\nu \bpm
    \tilde\nu_e\\ \tilde\nu_\mu\\ \tilde\nu_\tau\epm \ ,
\esp\ee
where the flavor-eigenstates are denoted explicitly and where the
physical eigenstates are mass-ordered, from the lightest to the heaviest. Again,
the four rotation matrices are implemented after splitting their real and
imaginary parts following the SLHA-2 conventions. The physical scalar fields
are implemented following the synthax of Ref.\ \cite{Christensen:2008py} and
contains all the information needed by the various \feynrules\ interfaces,
including the field redefinitions in the {\tt Definitions}
option. As an example, the redefinition of the scalar component of the
superfield $U_R^i$ is given by 
\begin{verbatim}
   Definitions -> { 
     URs[ff_, cc_] :> Module[{ff2}, subar[ff2,cc]*RuR[ff2,ff]]
   } , 
\end{verbatim}
where {\tt RuR} refers to the three last columns of the mixing matrix $R^u$ and
{\tt su} denote up-type squark mass-eigenstates.

Finally, we define Dirac representations $\psi$ for the fermions, because most generators work in terms of Dirac fermion rather than Weyl fermions, the former
being required at the Monte Carlo generator level. In terms of the Weyl fermions
introduced above, the Dirac fermions read
\be \bsp
  &\
    \psi_{u^i} = \bpm u_L^i \\ \bar u_{Ri}^c\epm \ , \quad 
    \psi_{d^i} = \bpm u_L^i \\ \bar d_{Ri}^c\epm \ , \quad 
    \psi_{e^i} = \bpm e_L^i \\ \bar e_{Ri}^c\epm \ , \quad 
    \psi_{\nu^i} = \bpm \nu_L^i \\ \bar \nu_{Ri}^c\epm \ , \\
  &\qquad \qquad
    \psi_{\chi^0_i} = \bpm \chi^0_i \\ \bar \chi^{0i} \epm \ , \quad 
    \psi_{\chi^\pm_i} = \bpm \chi^\pm_i \\ \bar \chi^{\mp i} \epm \ , \quad 
    \psi_{\tilde g} = \bpm i \tilde g \\ -i \overline {\tilde g}\epm \ .
\esp\ee 
The four-component Dirac and Majorana fermions are implemented following the
synthax of Ref.\ \cite{Christensen:2008py}, where we associate a four-component
fermion to its Weyl components through the {\tt WeylComponents} option of the
particle class \cite{Butterworth:2010ym}. As an example, the charged lepton
is implemented as 
\begin{verbatim}
   F[1] == { 
     ClassName->l, 
     SelfConjugate->False, 
     Indices->{Index[GEN]}, 
     FlavorIndex->GEN, 
     WeylComponents->{eLw,ERwbar}, 
     ...
   }, 
\end{verbatim}
where the dots stand for additional options such as those required by the Monte
Carlo tools.

The \feynrules\ function {\tt WeylToDirac} allows to perform the expansion of
the Weyl fermions in terms of the Dirac fields. However, we first need to
address the issue related to the antifundamental color representation which
the right-handed quark fields lie in. This is mandatory in order to have
one single color representation for the Dirac fields, \ie, the fundamental one,
and for the interfaces to the various tools linked to \feynrules\ to work
properly. Denoting $T$ and $\bar T$ the fundamental and antifundamental color
representations, and using the property $\bar T = - T^t$, the problem is solved
by implementing the instructions,
\begin{verbatim}
  Colourb = Colour  ,
  Lag = Lag /. { Tb[a_,i_,j_]->-T[a,j,i] } ,
\end{verbatim}
where {\tt Lag} denotes the MSSM Lagrangian. Then, since Dirac fermions are
defined only for physical particles, we start with an expansion of the $SU(2)_L$
multiplets in terms of their components,
\begin{verbatim}
  Lag = ExpandIndices[ Lag , FlavorExpand -> {SU2W, SU2D} ]  ,
\end{verbatim}
before eliminating the Weyl fermions from the Lagrangian,
\begin{verbatim}
  Lag = WeylToDirac[ Lag ] .
\end{verbatim}
The obtained Lagrangian is now suitable for the calculation of the Feynman rules
through the function {\tt FeynmanRules} or to be exported to \feynarts\ with the
function {\tt WriteFeynArtsOutput} or to any Monte Carlo tools linked to
\feynrules\ via the functions {\tt WriteCHOutput}, {\tt WriteMGOutput}, {\tt
WriteSHOutput}, {\tt WriteUFO} \cite{ufo} or {\tt WriteWOOutput}
\cite{Christensen:2010wz}.

\subsection{Validation} 

Our implementation has been validated against the (public) MSSM implementation
in the current \feynrules\ version 1.4.x. The latter is based on component
fields and Dirac fermions and has been validated
both against the literature and the built-in MSSM implementations in various
Monte Carlo tools \cite{Christensen:2009jx}. We have checked analytically that the
Feynman rules obtained with \feynrules\ using the superfield MSSM implementation
correspond to those found in the literature and verified that
\begin{verbatim}
   FeynmanRules[ LagSF - LagComponents ]
\end{verbatim}
provides an empty list of Feynman rules, ensuring that the two Lagrangians are
equal. In the expression above, {\tt LagSF} denotes the Lagrangian obtained
using the superspace module of \feynrules\ and {\tt LagComponents} is the one
available in the public version of the model.

\subsection{Gauge choice}
In Eq.\ \eqref{eq:HiggsMixing}, we have introduced the Goldstone bosons absorbed
by the weak gauge bosons to get their longitudinal polarization. Although absent
in calculations performed in unitarity gauge, they must be included if another gauge is used. 
In addition, a gauge-fixing Lagrangian must be included in the model, as well as ghost
fields associated to the gauge bosons, together with the corresponding Lagrangian
constructed from the BRST formalism. These two Lagrangians are included in
our MSSM implementation in the specific choice of the Feynman gauge.

The gauge-fixing Lagrangian is most conveniently derived after rewriting the
Higgs doublets in terms of eight real scalar fields $\phi_q^i$ with
$i=1,\ldots,4$. The Higgs doublets become
\be \bsp
  H_u =&\ \frac{1}{\sqrt{2}} 
    \bpm \phi_u^1 + i \phi_u^2 \\ \phi_u^3 + i \phi_u^4 \epm \equiv 
    \big(\phi_u^1, \phi_u^2, \phi_u^3, \phi_u^4, 0,0,0,0\big) \ , \\
  H_d =&\ \frac{1}{\sqrt{2}} 
    \bpm \phi_d^1 + i \phi_d^2 \\ \phi_d^3 + i \phi_d^4 \epm \equiv 
    \big(0,0,0,0, \phi_d^1, \phi_d^2, \phi_d^3, \phi_u^4 \big) \ ,
\esp\ee
and the gauge-fixing Lagrangian is given by
\be
  \lag_{\rm GF} = -\frac12 \Big(  G_B^\dag G_B + G_{Wk}^\dag G_W^k  +
    G_{ga}^\dag G_g^a \Big)
\label{eq:laggf}\ee 
with
\be\bsp
  G_B   =&\ \del_\mu B^\mu - g^\prime \Big(-\frac{i}{2} \langle H_u \rangle
    \Big) \cdot H_u - g^\prime \Big( \frac{i}{2} \langle H_d \rangle
    \Big) \cdot H_d  \ , \\
  G_W^k =&\ \del^\mu W_\mu^k - g_w \Big(-\frac{i}{2} \langle \sigma^k H_u \rangle
    \Big) \cdot H_u - g_w \Big( \frac{i}{2} \langle \sigma_k H_d \rangle
    \Big) \cdot H_d \ , \\
  G_g^a =&\ \del^\mu g_\mu^a \ ,
\esp\ee
where the dot product stands for the scalar product of the field space
introduced above. Comparing with Eq.\ \eqref{eq:vev}, we derive the only non-zero vevs
as $\langle \phi_u^3 \rangle=v_u$ and $\langle \phi_d^1 \rangle=v_d$.

The Fadeev-Popov ghost Lagrangian is derived from the gauge variation of the
gauge-fixing functions and reads, for the $B$-ghost $u_B$, the $W$-ghost $u_W$
and the gluon ghost $u_g$,
\be 
  \lag_{\rm ghost} = 
       -\bar u_B \del_\mu \del^\mu  u_B -
       \bar u_W \del_\mu D^\mu u_W -
       \bar u_g \del_\mu D^\mu u_g +
       \lag_{{\rm ghost},\phi} \ , 
\label{eq:laggh} \ee
where the adjoint representation indices are understood and the covariant
derivatives are taken in the adjoint representation,
\be
  D_\mu u_W^i = \partial_\mu u_W^i + g_w \epsilon^i{}_{jk}  W_\mu^j u_W^k \ ,
  \quad 
  D_\mu u_g^a = \partial_\mu u_g^a + g_s f^a{}_{bc} g_\mu^b u_g^c \ .
\ee
Finally, the scalar piece of the ghost Lagrangian is given by
\be \bsp 
  \lag_{{\rm ghost},\phi} = &\
    \bar u_A \Big[ 
      \big(T^A \langle H_u \rangle\big) \cdot 
        \big(T^{A^\prime} [\langle H_u \rangle + H_u]\big) + \\
   &\quad \qquad
      \big(T^A \langle H_d \rangle\big) \cdot 
        \big(T^{A^\prime} [\langle H_d \rangle + H_d]\big) \Big] u_{A^\prime}\ , 
\esp \ee
with $u_A = u_B$, $u_W^1$, $u_W^2$, $u_W^3$, the corresponding real
representation matrices being $T^A=-i Y$, $-i \sigma^1/2$, $-i \sigma^2/2$, $-i
\sigma^3/2$ and the dot product stands for the scalar product defined in the
basis $\{\phi_u^i, \phi_d^i\}$ introduced above. The ghosts related to the physical
gauge bosons are obtained by rotating those related to the gauge-eigenstates
parallel to the latter.

The MSSM implementation in \feynrules\ contains the two Lagrangians of Eq.\
\eqref{eq:laggf} and Eq.\ \eqref{eq:laggh} and is hence fully expressed in
Feynman gauge, the unitarity gauge being recovered by removing all the ghosts
and Golstone bosons from the Lagrangian.

\section{Conclusions}
In this paper we presented a superspace module for the \feynrules\ package, and the whole module 
is distributed together with the the \feynrules\ package starting from version 1.6.x.
The module allows to perform computations of superspace quantities involving Grassmann 
variables and is well suited for the implementation of Lagrangians in terms of superfields.
The package hence allows to implement supersymmetric models into \feynrules\ directly in terms
of superfields, the superfield expressions can then be expanded into component
fields and the corresponding coefficients of the Grassmann variables extracted
in an automated way. Furthermore, since the only piece of a supersymmetric
Lagrangian that is not fixed by supersymmetry and gauge invariance is the
superpotential, the package allows for an automatic generation of all the
kinetic terms and gauge interaction terms for the superfields, thus reducing the
implementation of a supersymmetric Lagrangian to the almost trivial task of
writing down the superpotential.

\section*{Acknowledgments}
The authors are grateful to N.\ Christensen, T.\ Hahn, F.\ Maltoni and M.\ Rausch de Traubenberg
for useful and inspiring discussions. BF acknowledges support by the
Theory-LHC France-initiative of the CNRS/IN2P3.

\bibliographystyle{elsarticle-num}
\bibliography{biblio}

\end{document}